\documentclass[american,english,graphicx,showpacs,preprint,acs]{revtex4}
\usepackage{ae,aecompl}
\usepackage[T1]{fontenc}
\usepackage[latin1]{inputenc}
\usepackage[letterpaper]{geometry}
\usepackage{amstext}
\usepackage{amssymb}

\makeatletter

\providecommand{\tabularnewline}{\\}

\@ifundefined{textcolor}{}
{%
 \definecolor{BLACK}{gray}{0}
 \definecolor{WHITE}{gray}{1}
 \definecolor{RED}{rgb}{1,0,0}
 \definecolor{GREEN}{rgb}{0,1,0}
 \definecolor{BLUE}{rgb}{0,0,1}
 \definecolor{CYAN}{cmyk}{1,0,0,0}
 \definecolor{MAGENTA}{cmyk}{0,1,0,0}
 \definecolor{YELLOW}{cmyk}{0,0,1,0}
 }

\makeatletter

\usepackage{float}



\makeatother

\usepackage{babel}

\makeatother

\usepackage{babel}
\begin{document}

\title{Extension of many-electron theory and approximate density functionals
to fractional charges and fractional spins}

\author{Weitao Yang}

\address{Department of Chemistry and Physics, Duke University, Durham, North
Carolina 27708, USA }

\author{Paula Mori-Sánchez}

\address{Departamento de Química, Universidad Autónoma de Madrid, Madrid 28049,
Spain}

\author{Aron J. Cohen}

\address{Department of Chemistry, University of Cambridge, Lensfield Road,
Cambridge, CB2 1EW, UK}
\begin{abstract}
The exact conditions for density functionals and density matrix functionals
in terms of fractional charges and fractional spins are known, and
their violation in commonly used functionals has been shown to be
the root of many major failures in practical applications. However,
approximate functionals are not normally expressed in terms of the
fractional variables. Here we develop a general framework for extending
approximate density functionals and many-electron theory to fractional-charge
and fractional-spin systems. Our development allows for the fractional
extension of any approximate theory that is a functional of $G^{0}$,
the one-electron Green's function of the non-interacting reference
system. The extension to fractional charge and fractional spin systems
is based on the ensemble average of the basic variable, $G^{0}$.
We demonstrate the fractional extension for the following theories:
(1) any explicit functional of the one-electron density, such as the
local density approximation and generalized gradient approximations;
(2) any explicit functional of the one-electron density matrix of
the non-interacting reference system, such as the exact exchange functional
(or Hartree-Fock theory) and hybrid functionals; (3) many-body perturbation
theory; and (4) random-phase approximations. A general rule for such
an extension has also been derived through scaling the orbitals and
should be useful for functionals where the link to the Green's function
is not obvious. The development thus enables the examination of approximate
theories against known exact conditions on the fractional variables
and the analysis of their failures in chemical and physical applications
in terms of violations of exact conditions of the energy functionals.
The present work should facilitate the calculation of chemical potentials
and fundamental band gaps with approximate functionals and many-electron
theories through the energy derivatives with respect to the fractional
charge. It should play an important role in developing accurate approximate
density functionals and many-body theory.
\end{abstract}

\date{\today{}}

\maketitle

\section{Introduction}

Exact conditions for the ground-state energy of many-electron systems
play a very important role in the development and understanding of
approximate density functional and many body theories. Particularly
relevant are the exact conditions for fractional charges \cite{Perdew821691}
and fractional spins \cite{Cohen08121104,Mori-Sanchez0966403}, that
have highlighted some key failures of approximate density functionals
which are connected to major failures in chemical and physical applications
\cite{Cohen08792} associated with the dissociation of molecular ions,
polarizabilities, barrier heights, magnetic properties, fundamental
band-gaps and strongly-correlated systems.

The main problem of DFT to give accurate energy gaps in finite and
bulk systems can be traced to the delocalization error, which is defined
as the deviation of a given approximate functional from the exact
linear behavior in fractional charges\cite{Mori-Sanchez08146401}.
Consequently, most approximate functionals tend to over-delocalize
the added electron or hole or give unphysically low energies for delocalized
electrons. . The consequences of the delocalization error can be seen
not only in the prediction of derivative properties such as band gaps
and charge transfer excitations\cite{Dreuw044007}, or the energy
of stretched molecular ions \cite{Zhang982604}, but also in thermochemistry
and structures of molecules at equilibrium \cite{Johnson08204112,Heaton-Burgess10234113}.
Progress has been made in designing functionals with reduced delocalization
error \cite{Cohen07191109,Baer1085,Tsuneda10174101,Zheng1126403,xxx}.
Because of the delocalization error, currently, reliable band gap
prediction is dependent on the use of many-body Green's function theory,
such as the GW approximation \foreignlanguage{american}{\cite{Onida02601}}.

Analogously, it has been proven \cite{Cohen08121104} that fractional
spins arise in systems with large static correlation energy (strongly
correlated systems). Static correlation error of approximate functionals
is defined as the deviation in the energy of fractional-spin states
from the constancy condition defined by the energy of the comprising
pure spin states. This leads to another set of different failures
in DFT such as incorrect chemical bond dissociation or failure for
the band structure prediction of Mott insulators \cite{Cohen08121104,Mori-Sanchez0966403}.

The simple physical picture of fractional charges and fractional spins
comes from molecular dissociation \cite{Yang005172}. For example,
fractional-charge hydrogen atoms result from the dissociation of H$_{2}^{+}$
\cite{Zhang982604}, and fractional-spin hydrogen atoms from the dissociation
of H$_{2}$ \cite{Cohen08121104}. This is key as it makes possible
the direct numerical verification for the fractional extensions of
approximate theories: The fractional-charge or fractional-spin calculations
have to yield energies that agree with the limits of the corresponding
molecular dissociation \cite{Cohen12289}.

For the extension of approximate functionals of the density or the
first-order density matrix to fractional charges, $N+\delta$ $(0<\delta<1)$,
the ensemble average of the density \cite{Perdew821691} or the one-electron
density matrix \cite{Yang005172} can be used
\begin{equation}
\rho_{1}(N+\delta)=(1-\delta){\rho_{1}}(N)+\delta{\rho_{1}}(N+1).\label{eq:rho1}
\end{equation}
This leads to the simple extension of fractional occupation of the
orbitals, $\rho_{1}({\bf r},{\bf r}^{\prime})=\sum_{i}n_{i}\phi_{i}({\bf r})\phi_{i}({\bf r}^{\prime})$,
which has enabled the extension of methods such as local density approximation
(LDA), generalized gradient approximation (GGA) or Hartree-Fock (HF)
to fractional charge and fractional spins \cite{Mori-Sanchez06201102,Perdew0740501,Cohen08121104}.

These exact conditions for fractional charges and fractional spins
are just as important for many-body theory based methods, where an
extension to fractional occupations is still required. However, this
generalization is not always a trivial question because the basic
variable is no longer the density or the one-particle density matrix.
Furthermore, the extension to fractional occupations does not correspond
to a finite temperature formulation of the theory evaluated at T=0
to linearly combine the energies in the traditional ensemble perspective.
For M\o ller Plesset theory (MP2) and the random phase approximation
(RPA)\cite{Negele88} \cite{Ren127447,Eshuis121084}, which depend
on the unoccupied orbitals and eigenvalues, the basic variable is
the one-electron Green's function, which now plays the same role as
the one-electron density matrix for approximate density functionals.
The fractional extensions have been made for the M\o ller Plesset
theory (MP2) \cite{Cohen09786} and for the random phase approximation
\cite{Mori-Sanchez09,Mori-Sanchez1242507}, but without rigorous derivation.

In this work, we will show that the basic variable is an ensemble
average of the one-electron Green's function for the non-interacting
systems. As it will be shown, this result agrees with the Eq. (\ref{eq:rho1})
and allows the correct extension of many-body methods to fractional
charges and fractional spins.

\section{Fractional Charges and Fractional Spins Based on One-electron Green's
Functions}

There is a large class of approximate functionals or many-electron
theories, which can be cast as functionals of the one-electron Green's
function of the non-interacting reference system. Because the electron
density and the first-order density matrix for the non interacting
reference system is given by the one-electron Green's function, thus
common functionals of the density or the one-electron density matrix
are included.

The one-electron Green's function of an $N$-electron system in its
ground state $\Big|\Psi_{0}^{N}\Big\rangle$ is defined as
\begin{eqnarray}
G^{N}(i,j;t-t') & = & -i\left\langle \Psi_{0}^{N}\left|T(a_{i}(t)a_{j}^{\dagger}(t'))\right|\Psi_{0}^{N}\right\rangle ,\label{green1}
\end{eqnarray}
 in terms of the time ordering operator $T$ and the creation $a^{\dagger}$
and annihilation $a$ operators. Note that the index $i$ includes
spin. In terms of combined spatial and spin coordinates $\mathbf{x}$
\begin{eqnarray}
G^{N}(\mathbf{x},\mathbf{x}';t-t') & = & -i\left\langle \Psi_{0}^{N}\left|T(\hat{\psi}(\mathbf{x},t)\hat{\psi}^{\dagger}(\mathbf{x}',t'))\right|\Psi_{0}^{N}\right\rangle \nonumber \\
 & = & \sum_{ij}G^{N}(i,j;t-t')\phi_{i}(\mathbf{x})\phi_{j}^{*}(\mathbf{x}'),
\end{eqnarray}
 with the field operators,
\begin{equation}
\hat{\psi}(\mathbf{x},t)=\sum_{i}\phi_{i}(\mathbf{x})a_{i}(t)
\end{equation}

\begin{equation}
\hat{\psi}^{\dagger}(\mathbf{x},t)=\sum_{i}\phi_{i}^{*}(\mathbf{x})a_{i}^{\dagger}(t)
\end{equation}
expressed as linear combinations of creation and annihilation operators
where the coefficients are the single particle spin orbitals $\left\{ \left|\phi_{i}\right\rangle \right\} $
and the sums run over all possible single particle states.

The one-electron Green's function can also be expressed in terms of
a complete set of eigenstates $\left\{ \Big|\Psi_{m}^{N+1}\Big\rangle\right\} $
and eigenvalues $\left\{ E_{m}^{N+1}\right\} $ of the system Hamiltonian:
\begin{eqnarray}
G^{N}(i,j;t-t') & = & -i\left\{ \theta(t-t')\sum_{m}e^{i(E_{0}^{N}-E_{m}^{N+1})(t-t')}\langle\Psi_{0}^{N}\Big|a_{i}\Big|\Psi_{m}^{N+1}\Big\rangle\Big\langle\Psi_{m}^{N+1}\Big|a_{j}^{\dagger}\Big|\Psi_{0}^{N}\Big\rangle\right.\nonumber \\
 &  & \left.-\theta(t'-t)\sum_{n}e^{i(E_{0}^{N}-E_{n}^{N-1})(t'-t)}\Big\langle\Psi_{0}^{N}\Big|a_{j}^{\dagger}\Big|\Psi_{n}^{N-1}\Big\rangle\Big\langle\Psi_{n}^{N-1}\Big|a_{i}\Big|\Psi_{0}^{N}\Big\rangle\right\} ,
\end{eqnarray}
which leads to the Lehmann representation in energy, given by the
Fourier transform
\begin{eqnarray}
G^{N}(i,j;E) & = & \int_{-\infty}^{+\infty}d(t-t')e^{iE(t-t')}G^{N}(i,j;t-t')\nonumber \\
 & = & \sum_{m}\frac{\Big\langle\Psi_{0}^{N}\Big|a_{i}\Big|\Psi_{m}^{N+1}\Big\rangle\Big\langle\Psi_{m}^{N+1}\Big|a_{j}^{\dagger}\Big|\Psi_{0}^{N}\Big\rangle}{E-(E_{m}^{N+1}-E_{0}^{N})+i\eta}+\sum_{n}\frac{\Big\langle\Psi_{0}^{N}\Big|a_{j}^{\dagger}\Big|\Psi_{n}^{N-1}\Big\rangle\Big\langle\Psi_{n}^{N-1}\Big|a_{i}\Big|\Psi_{0}^{N}\Big\rangle}{E-(E_{0}^{N}-E_{n}^{N-1})-i\eta}
\end{eqnarray}

For a non-interacting system that is described by a normalized Slater
determinant $\Phi_{0}^{N}$ with one-electron orbitals $\left\{ \left|\phi_{i}\right\rangle \right\} $
and orbital energies $\left\{ \varepsilon_{i}\right\} $, its single-particle
Green's function is given by
\begin{eqnarray}
G^{0,N}(i,j;t-t') & = & -i\left\langle \Phi_{0}^{N}\left|T(a_{i}(t)a_{j}^{\dagger}(t'))\right|\Phi_{0}^{N}\right\rangle \nonumber \\
 & = & -i\delta_{ij}e^{-i\varepsilon_{i}(t-t')}\left\{ \theta(t-t')\theta(i-F)-\theta(t'-t)\theta(F-i)\right\} ,
\end{eqnarray}
 where $F$ is a number larger than the index for the highest occupied
orbital but smaller than the index for the lowest unoccupied orbital.
In coordinate and spin space
\begin{eqnarray}
G^{0,N}(\mathbf{x},\mathbf{x}';t-t') & = & -i\left\langle \Phi_{0}^{N}\left|T(\hat{\psi}(\mathbf{x},t)\hat{\psi}^{\dagger}(\mathbf{x}',t'))\right|\Phi_{0}^{N}\right\rangle \\
 & = & \sum_{ij}G^{0,N}(i,j;t-t')\phi_{i}(\mathbf{x})\phi_{j}^{*}(\mathbf{x}').
\end{eqnarray}
In energy representation,
\begin{eqnarray}
G^{0,N}(i,j;E) & = & \int_{-\infty}^{+\infty}d(t-t')e^{iE(t-t')}G^{0,N}(i,j;t-t')\\
 & = & \delta_{ij}\left\{ \frac{\theta(i-F)}{E-\varepsilon_{i}+i\eta}+\frac{\theta(F-i)}{E-\varepsilon_{i}-i\eta}\right\} ,
\end{eqnarray}
and
\begin{eqnarray}
G^{0,N}(\mathbf{x},\mathbf{x}';E) & = & \sum_{ij}G^{0,N}(i,j;E)\phi_{i}(\mathbf{x})\phi_{j}^{*}(\mathbf{x}')\nonumber \\
 & = & \sum_{i>F}\frac{1}{E-\varepsilon_{i}+i\eta}\phi_{i}(\mathbf{x})\phi_{i}(\mathbf{x}')^{*}+\sum_{i<F}\frac{1}{E-\varepsilon_{i}-i\eta}\phi_{i}(\mathbf{x})\phi_{i}(\mathbf{x}')^{*}\nonumber \\
 & = & \sum_{a}\frac{1}{E-\varepsilon_{a}+i\eta}\phi_{a}(\mathbf{x})\phi_{a}(\mathbf{x}')^{*}+\sum_{i}\frac{1}{E-\varepsilon_{i}-i\eta}\phi_{i}(\mathbf{x})\phi_{i}(\mathbf{x}')^{*}
\end{eqnarray}
where $a,b,c,d$ are particle indexes (unoccupied), $i,j,k,l$ are
hole indexes (occupied) and $m,n,o,p$ are general indexes. Sometimes,
$i,j,k,l$ are used for general indexes, in specific cases as in Eqs.
(\ref{green1}).

To make the extension to fractional charge and fractional spin systems,
we construct an ensemble of systems that are described with the \emph{same}
non-interacting reference Hamiltonian. For a fractionally charged
system with $N+\delta$ electrons, where the spin character of the
additional fractional charge $\delta$ is expressed through the orbital
and its corresponding occupation number $n_{i}$, we define the single-particle
Green's function as the following ensemble average
\begin{eqnarray}
G^{0,N+\delta}(i,j;t-t') & = & (1-\delta)G^{0,N}(i,j;t-t')+\delta G^{0,N+1}(i,j;t-t')\nonumber \\
 & = & -i\delta_{ij}e^{-i\varepsilon_{i}(t-t')}\left\{ \theta(t-t')(1-n_{i})-\theta(t'-t)n_{i}\right\} \label{eq:G_0_t_frac}
\end{eqnarray}
 or
\begin{eqnarray}
G^{0,N+\delta}(i,j;E) & = & (1-\delta)G^{0,N}(i,j;E)+\delta G^{0,N+1}(i,j;E)\nonumber \\
 & = & \delta_{ij}\left\{ \frac{(1-n_{i})}{E-\varepsilon_{i}+i\eta}+\frac{n_{i}}{E-\varepsilon_{i}-i\eta}\right\} .\label{eq:G_0_E_frac}
\end{eqnarray}
 where
\begin{equation}
\sum_{i}n_{i}=N+\delta
\end{equation}
 In coordinate and spin space,
\begin{eqnarray}
G^{0,N+\delta}(\mathbf{x},\mathbf{x}';E) & = & \sum_{ij}G^{0,N+\delta}(i,j;E)\phi_{i}(\mathbf{x})\phi_{j}^{*}(\mathbf{x}')\nonumber \\
 & = & \sum_{i\geq F}\frac{(1-n_{i})}{E-\varepsilon_{i}+i\eta}\phi_{i}(\mathbf{x})\phi_{i}(\mathbf{x}')^{*}+\sum_{i\leq F}\frac{n_{i}}{E-\varepsilon_{i}-i\eta}\phi_{i}(\mathbf{x})\phi_{i}(\mathbf{x}')^{*}\nonumber \\
 & = & \sum_{a}\frac{1}{E-\varepsilon_{a}+i\eta}\tilde{\phi}_{a}(\mathbf{x})\tilde{\phi}_{a}(\mathbf{x}')^{*}+\sum_{i}\frac{1}{E-\varepsilon_{i}-i\eta}\tilde{\phi}_{i}(\mathbf{x})\tilde{\phi}_{i}(\mathbf{x}')^{*}\label{eq:Green(NpDelta)}
\end{eqnarray}
 Fractional occupations occur only at the frontier levels, because
we require the non-interacting ground state representation of an interacting
system.

We have \emph{extended} the index $a$ for unoccupied orbital (particle)
to include fractionally unoccupied states, and index $i$ for occupied
states (hole) to include fractionally occupied states. We also introduce
the occupation-scaled unoccupied orbitals

\begin{equation}
\tilde{\phi}_{a}(\mathbf{x}_{1})=\sqrt{1-n_{a}}\phi_{a}(\mathbf{x}_{1})\label{eq:scaled_u}
\end{equation}
and the occupation-scaled occupied orbitals

\begin{equation}
\tilde{\phi}_{i}(\mathbf{x}{}_{2})=\sqrt{n_{i}}\phi_{i}(\mathbf{x}{}_{2}).\label{eq:scaled_o}
\end{equation}
Our extension allows the direct incorporation of fractional charges
and spins into many-body theories based on Green's functions.

The expression $G^{0,N+\delta}(\mathbf{x},\mathbf{x}';E)$ Eq. (\ref{eq:Green(NpDelta)}),
is the key result, which underlies \emph{a simple rule} for extending
approximate density functionals and many-body theories to fractional
charges and fractional spins: Notice that $G^{0,N+\delta}(\mathbf{x},\mathbf{x}';E)$
has the same form as $G^{0,N}(\mathbf{x},\mathbf{x}';E)$, except
that i) the parent orbitals need to be replaced by the occupation-scaled
orbitals, ii) the set of occupied orbitals includes fractionally occupied
ones, and iii) the set of unoccupied orbitals includes the fractionally
unoccupied ones. Thus fractional orbitals enter into both sets: as
fractionally occupied, and as fractionally unoccupied.

Because the fractional orbitals enter into the formalism as fractional
hole/occupied orbital and also as fractional particle/unoccupied orbital,
the matrix representation for $G^{0,N+\delta}(i,j;E)$ can be written
in the extended matrix of $(n_{o}+n_{f}+n_{f}+n_{u})$, where $n_{o}$
is the number of (fully) occupied orbitals, $n_{f}$ is the number
of fractional orbitals, and $n_{u}$ is the number of fully unoccupied
orbitals. $G^{0,N+\delta}(i,j;E)$ is a diagonal matrix with four
blocks of states: occupied states of the size $n_{o}$ , fractional
occupied state of the size $n_{f}$, fractional unoccupied states
of the size $n_{f}$, and unoccupied states of the size $n_{u}$.
The structure is
\begin{equation}
\mathbf{G}^{0,N+\delta}=\left(\begin{array}{cccc}
\frac{1}{E-\varepsilon_{i}-i\eta}\\
 & \frac{n_{i}}{E-\varepsilon_{i}-i\eta}\\
 &  & \frac{(1-n_{i})}{E-\varepsilon_{i}+i\eta}\\
 &  &  & \frac{1}{E-\varepsilon_{i}+i\eta}
\end{array}\right).\label{eq:Green_Matrix_frac}
\end{equation}
 Note that in Eq.(\ref{eq:Green_Matrix_frac}), the Green's function
has been unfolded into a matrix with the dimension of $(n_{o}+n_{f}+n_{f}+n_{u})$,
larger than $(n_{o}+n_{f}+n_{u})$, which is the dimension of all
one-electron states. In this way, the matrix elements in orbital space
in Eq.(\ref{eq:Green_Matrix_frac}) are different from those in Eq.(\ref{eq:G_0_E_frac}),
but they have the same real-space representation of Eq. (\ref{eq:Green(NpDelta)}).

In the following, we apply the fractional extension $G^{0,N+\delta}$
in various approximations and work out the details of the extension
to various types of approximate functionals in many-body theory. Remarkably,
in all the cases the simple extension rule applies.

\section{Hartree Fock and Perturbation Theory}

\subsection{The density and density matrix from one-particle Green's functions}

The non-interacting one-electron density matrix, corresponding the
hole part in the many-body language, is given by
\begin{eqnarray}
\rho_{s}(i,j) & = & \int_{-\infty}^{+\infty}\frac{dE}{2\pi i}e^{iE\eta}G^{0,N+\delta}(i,j;E)\nonumber \\
 & = & \int_{C\uparrow}\frac{dE}{2\pi i}G^{0,N+\delta}(i,j;E)\nonumber \\
 & = & -iG^{0,N+\delta}(i,j;t,t^{+})\nonumber \\
 & = & \delta_{ij}n_{i},
\end{eqnarray}
where the integration along the path $C\uparrow$ is the integration
from $-\infty$ to $+\infty$ and closed on the negative side of the
complex $E$ plane. Analogously the non-interacting particle matrix
of many body-theory (the virtual state density) is
\begin{eqnarray}
\bar{\rho}_{s}(i,j) & = & \int_{-\infty}^{+\infty}\frac{dE}{2\pi i}e^{-iE\eta}G^{0,N+\delta}(i,j;E)\nonumber \\
 & = & iG^{0,N+\delta}(i,j;t^{+}-t)\nonumber \\
 & = & \delta_{ij}(1-n_{i}).
\end{eqnarray}

In coordinate and spin space
\begin{eqnarray}
\rho_{s}(\mathbf{x},\mathbf{x}') & = & \sum_{ij}\rho_{s}(i,j)\phi_{i}(\mathbf{x})\phi_{j}^{*}(\mathbf{x}')\nonumber \\
 & = & -iG^{0,N+\delta}(\mathbf{x},\mathbf{x}';t,t^{+})\nonumber \\
 & = & \sum_{i}n_{i}\phi_{i}(\mathbf{x})\phi_{i}^{*}(\mathbf{x}'),\label{eq:DensityMatrix}
\end{eqnarray}
and
\begin{eqnarray}
\bar{\rho}_{s}(\mathbf{x},\mathbf{x}') & = & \sum_{ij}\bar{\rho}_{i,j}\phi_{i}(\mathbf{x})\phi_{j}^{*}(\mathbf{x}')\nonumber \\
 & = & \sum_{i}(1-n_{i})\phi_{i}(\mathbf{x})\phi_{i}^{*}(\mathbf{x}'),
\end{eqnarray}
 such that
\begin{equation}
\rho_{s}(\mathbf{x},\mathbf{x}')+\bar{\rho}_{s}(\mathbf{x},\mathbf{x}')=\delta(\mathbf{x}-\mathbf{x}').
\end{equation}
The electron density can thus be expressed in terms of the occupation
numbers as
\begin{eqnarray}
\rho(\mathbf{x}) & = & \sum_{i}n_{i}\phi_{i}(\mathbf{x})\phi_{i}^{*}(\mathbf{x})\label{eq:Density}
\end{eqnarray}

This equation is consistent with previous work \cite{Mori-Sanchez06201102,Cohen08121104,Mori-Sanchez0966403}
for any functional of $\rho(\mathbf{x})$ or $\rho_{s}(\mathbf{x},\mathbf{x}')$
such as LDA, GGA or HF. Here and hereafter, we suppress the superscripts
{}``$N+\delta$'' for the density and particle matrix for the non-interacting
fractional reference systems, when the fractional context is unambiguous.

\subsection{Perturbation theory based on one-particle Green's functions }


We will derive energy functionals of $G^{0,N+\delta}$ within the
perturbation theory. The many-electron Hamiltonian is given by
\begin{equation}
H=\sum_{i}\left(-\frac{1}{2}\nabla_{i}^{2}+v_{ext}(\mathbf{r}_{i})\right)+\sum_{i\neq j}v(r_{ij}),
\end{equation}
 or equivalently in Fock space,
\begin{eqnarray}
H & = & \sum_{ij}h_{ij}a_{i}^{\dagger}a_{j}+\frac{1}{2}\sum_{ijkl}\langle ij|kl\rangle a_{i}^{\dagger}a_{j}^{\dagger}a_{l}a_{k}\nonumber \\
 & = & \sum_{ij}h_{ij}a_{i}^{\dagger}a_{j}+\frac{1}{4}\sum_{ijkl}\langle ij||kl\rangle a_{i}^{\dagger}a_{j}^{\dagger}a_{l}a_{k},
\end{eqnarray}
 where
\begin{equation}
h_{ij}=\langle\phi_{i}|-\frac{1}{2}\nabla^{2}+v_{ext}(\mathbf{r})|\phi_{j}\rangle=\langle i|-\frac{1}{2}\nabla^{2}+v_{ext}(\mathbf{r})|j\rangle,
\end{equation}
 and
\begin{equation}
\langle ij|kl\rangle=\int\int d\mathbf{x}_{1}d\mathbf{x}_{2}\phi_{i}^{*}(\mathbf{x}_{1})\phi_{j}^{*}(\mathbf{x}_{2})v(r_{12})\phi_{k}(\mathbf{x}_{1})\phi_{l}(\mathbf{x}_{2}),
\end{equation}

\begin{eqnarray}
\langle ij||kl\rangle & = & \langle ij|kl\rangle-\langle ij|lk\rangle.
\end{eqnarray}

Consider a non-interacting reference system which can have a local
or non-local potential $v_{s}=v_{ext}+u$, of the form $v_{s}(\mathbf{x})$
or $v_{s}(\mathbf{x,x'})$ respectively, as determined by the nature
of $u$,
\begin{equation}
\left(-\frac{1}{2}\nabla^{2}+v_{ext}+u\right)\left|\phi_{i}\right\rangle =\varepsilon_{i}\left|\phi_{i}\right\rangle .\label{eq:Non-interactingRef}
\end{equation}
 The equation for the Green's function of the physical interacting
system is
\begin{equation}
\left[E-\left(h+u+\Sigma^{*}\right)\right]G(E)=I\label{greenequation}
\end{equation}
 and the corresponding Dyson equation
\begin{equation}
G(E)=G^{0}(E)+{\displaystyle G^{0}(E)\Sigma^{*}(E)G(E)}\label{dysonequation}
\end{equation}
 The irreducible self energy $\Sigma^{*}(E)$, expanded up to second-order
perturbation in the electron-electron interaction \cite{Dickhoff05},
is given by
\begin{equation}
\Sigma^{*}(k,l,E)=-\langle k|u|l\rangle+\lambda\Sigma^{*(1)}(k,l,E)+\lambda^{2}\Sigma^{*(2)}(k,l,E)\label{eq:SelfE-2Order}
\end{equation}
 where $\lambda$ is the order parameter representing the electron-electron
interaction, and the first- and second-order contributions are
\begin{eqnarray}
\Sigma^{*(1)}(i,j,E) & = & \int_{C\uparrow}\frac{dE'}{2\pi i}\sum_{kl}\langle ik||jl\rangle G^{0}(l,k;E')\\
 & = & \sum_{kl}\langle ik||jl\rangle\rho_{s}(l,k)\nonumber \\
 & = & \sum_{k}\langle ik||jk\rangle n_{k}\nonumber \\
 & = & J_{ij}-K_{ij},
\end{eqnarray}
 and
\begin{eqnarray}
\Sigma^{*(2)}(i,j,E) & = & -\frac{1}{2}\int_{-\infty}^{+\infty}\frac{dE_{1}}{2\pi i}\int_{-\infty}^{+\infty}\frac{dE_{2}}{2\pi i}\nonumber \\
 &  & \sum_{klm}\sum_{npq}\langle ik||lm\rangle\langle np||jq\rangle G^{0}(l,n;E_{1})G^{0}(m,p;E_{2})G^{0}(q,k;E_{1}+E_{2}-E)\nonumber \\
 & = & \frac{1}{2}\sum_{lmq}\langle iq||lm\rangle\langle lm||jq\rangle\left\{ \frac{(1-n_{l})(1-n_{m})n_{q}}{E-\varepsilon_{l}-\varepsilon_{m}+\varepsilon_{q}+i\eta}+\frac{n_{l}n_{m}(1-n_{q})}{E-\varepsilon_{l}-\varepsilon_{m}+\varepsilon_{q}-i\eta}\right\} .\label{eq:SelfEnergy2}
\end{eqnarray}
For the details of the integration leading to Eq. (\ref{eq:SelfEnergy2}),
see Section \ref{sub:The-second-order} of the Appendix.

Now we introduce the Hamiltonian $H(\lambda)$ as a function of the
coupling parameter $\lambda$
\begin{equation}
H({\lambda})=H_{0}+\lambda H_{1}\label{eq:AC_1}
\end{equation}
 where
\begin{eqnarray}
H_{0} & = & \sum_{ij}(h_{ij}+u_{ij})a_{i}^{\dagger}a_{j}\nonumber \\
 & = & \sum_{i}\varepsilon_{i}a_{i}^{\dagger}a_{i},
\end{eqnarray}
 and
\begin{eqnarray}
H_{1} & = & \frac{1}{2}\sum_{ijkl}\langle ij|kl\rangle a_{i}^{\dagger}a_{j}^{\dagger}a_{l}a_{k}-\sum_{ij}u_{ij}a_{i}^{\dagger}a_{j}.\nonumber \\
 & = & \frac{1}{2}\sum_{ijkl}\langle ij|kl\rangle a_{i}^{\dagger}a_{j}^{\dagger}a_{l}a_{k}-\sum_{ij}u_{ij}a_{i}^{\dagger}a_{j}.
\end{eqnarray}
Thus, $H(0)=H_{0}$, $H(1)=H_{0}+H_{1}=H$. Then the total energy
as a function of $\lambda$ is
\begin{eqnarray}
E(\lambda) & = & \left\langle \Psi_{0}^{N}(\lambda)\left|H({\lambda})\right|\Psi_{0}^{N}(\lambda)\right\rangle ,
\end{eqnarray}
and its derivative is
\begin{equation}
\frac{dE(\lambda)}{d\lambda}=\left\langle \Psi_{0}^{N}(\lambda)\left|H_{1}\right|\Psi_{0}^{N}(\lambda)\right\rangle .
\end{equation}
The total energy $E(1)$ for the physical system is given by

\begin{eqnarray}
 &  & E(1)-E(0)\nonumber \\
 & = & \intop_{0}^{1}d\lambda\left\langle \Psi_{0}^{N}(\lambda)\left|H_{1}\right|\Psi_{0}^{N}(\lambda)\right\rangle \nonumber \\
 & = & \intop_{0}^{1}d\lambda\left\{ -\sum_{ij}u_{ij}\left\langle \Psi_{0}^{N}(\lambda)\left|a_{i}^{\dagger}a_{j}\right|\Psi_{0}^{N}(\lambda)\right\rangle +\frac{1}{2}\sum_{ijkl}\langle ij|kl\rangle\left\langle \Psi_{0}^{N}(\lambda)\left|a_{i}^{\dagger}a_{j}^{\dagger}a_{l}a_{k}\right|\Psi_{0}^{N}(\lambda)\right\rangle \right\} ,\label{eq:TotalE0}
\end{eqnarray}
where we suppress the index of $\lambda$ in $a_{i}$. This equation
will lead to various useful expressions of the total energy in terms
of Green's functions, $\mathbf{G}^{0}(E)$ and $\mathbf{G}(E)$, and
the self energies.

Using the equation of motion, we obtain, with details given in Section
(\ref{sub:Use-of-theEOM}) of the Appendix, $E(1)$ in terms of $\mathbf{G}^{\lambda}(E)$
and the irreducible self energy $\mathbf{\Sigma}^{*\lambda}(E)$

\begin{equation}
E(1)-E(0)=\frac{1}{2}\intop_{0}^{1}d\lambda\int_{-\infty}^{+\infty}\frac{dE}{2\pi i}e^{iE\eta}\mathrm{Tr}\left(-\mathbf{u}+\frac{1}{\lambda}\mathbf{\Sigma}^{*\lambda}(E)\right)\mathbf{G}^{\lambda}(E).\label{eq:TotelE}
\end{equation}
Using the Dyson equation
\begin{equation}
\mathbf{G}^{\lambda}(E)=\mathbf{G}^{0}(E)+\mathbf{G}^{0}(E)\mathbf{\Sigma}^{\lambda}(E)\mathbf{G}^{0}(E),
\end{equation}
in terms of the reducible self-energy $\mathbf{\Sigma}^{\lambda}(E)$,
\begin{eqnarray}
\mathbf{\Sigma}^{\lambda}(E) & = & \mathbf{\Sigma}^{*\lambda}(E)+\mathbf{\Sigma}^{\lambda}(E)\mathbf{G\mathnormal{^{0}(E)}}\mathbf{\Sigma}^{*\lambda}(E)\nonumber \\
 & = & \mathbf{\Sigma}^{*\lambda}(E)+\mathbf{\Sigma}^{*\lambda}(E)\mathbf{G}^{0}(E)\mathbf{\Sigma}^{\lambda}(E)\label{eq:ReducibleSelfE}
\end{eqnarray}
we obtain $E(1)$ in terms of the reducible self-energy $\mathbf{\Sigma}^{\lambda}(E)$
and $\mathbf{G}^{0}(E)$:
\begin{eqnarray}
E(1)-E(0) & = & \frac{1}{2}\intop_{0}^{1}d\lambda\int_{-\infty}^{+\infty}\frac{dE}{2\pi i}e^{iE\eta}\mathrm{Tr}\left[\mathbf{-}\mathbf{u}\left(\mathbf{G}^{0}(E)+\mathbf{G}^{0}(E)\mathbf{\Sigma}^{\lambda}(E)\mathbf{G}^{0}(E)\right)+\frac{1}{\lambda}\mathbf{\Sigma}^{\lambda}(E)\mathbf{G}^{0}(E)\right]\label{eq:TotalE3}
\end{eqnarray}
We now use Eq. (\ref{eq:ReducibleSelfE}) and express $\mathbf{\Sigma}^{\lambda}(E)$
in terms of the irreducible self-energy $\mathbf{\Sigma}^{*\lambda}(E)$
and $\mathbf{G}^{0}(E)$:

\begin{eqnarray}
\mathbf{\Sigma}^{\lambda}(E) & = & \mathbf{\Sigma}^{*\lambda}(E)\left[1-\mathbf{G}^{0}(E)\mathbf{\Sigma}^{*\lambda}(E)\right]^{-1},\label{eq:ReducibleSelfE2}
\end{eqnarray}
which can be used to obtain the perturbation expansion for $\mathbf{\Sigma}^{\lambda}(E)$ to
input into Eq. (\ref{eq:TotalE3}) to obtain the total energy. In
addition, from Eq. (\ref{eq:TotelE}), and using $\left(\mathbf{G}^{\lambda}(E)\right)^{-1}=\left(\mathbf{G}^{0}(E)\right)^{-1}-\mathbf{\Sigma}^{*\lambda}(E)$,
we have

\begin{eqnarray}
E(1)-E(0) & = & \frac{1}{2}\intop_{0}^{1}d\lambda\int_{-\infty}^{+\infty}\frac{dE}{2\pi i}e^{iE\eta}\mathrm{Tr}\left(-\mathbf{u}+\frac{1}{\lambda}\mathbf{\Sigma}^{*\lambda}(E)\right)\left(\left(\mathbf{G}^{0}(E)\right)^{-1}-\mathbf{\Sigma}^{*\lambda}(E)\right)^{-1},\label{eq:TotelE2}
\end{eqnarray}
 which expresses the total energy directly in terms of the irreducible
self-energy $\mathbf{\Sigma^{*}}(E)$ and $\mathbf{G}^{0}(E)$.

To second order, using Eq.(\ref{eq:SelfE-2Order}) in Eq. (\ref{eq:ReducibleSelfE})
\begin{eqnarray}
\mathbf{\Sigma}^{\lambda}(E) & = & \left[\mathbf{-}\lambda\mathbf{u}+\lambda\mathbf{\Sigma}^{*(1)}(E)+\lambda^{2}\mathbf{\Sigma}^{*(2)}(E)\right]\nonumber \\
 &  & +\left[\mathbf{-}\lambda\mathbf{u}+\lambda\mathbf{\Sigma}^{*(1)}(E)+\lambda^{2}\mathbf{\Sigma}^{*(2)}(E)\right]\mathbf{G\mathnormal{^{0}(E)}}\left[\mathbf{-}\lambda\mathbf{u}+\lambda\mathbf{\Sigma}^{*(1)}(E)+\lambda^{2}\mathbf{\Sigma}^{*(2)}(E)\right]+...\nonumber \\
 & = & \left[\mathbf{-}\lambda\mathbf{u}+\lambda\mathbf{\Sigma}^{*(1)}(E)+\lambda^{2}\mathbf{\Sigma}^{*(2)}(E)\right]+\left[\mathbf{-}\lambda\mathbf{u}+\lambda\mathbf{\Sigma}^{*(1)}(E)\right]\mathbf{G\mathnormal{^{0}(E)}}\left[\mathbf{-}\lambda\mathbf{u}+\lambda\mathbf{\Sigma}^{*(1)}(E)\right]+...\nonumber \\
 & = & \lambda\left[\mathbf{-}\mathbf{u}+\mathbf{\Sigma}^{*(1)}(E)\right]+\lambda^{2}\left\{ \mathbf{\Sigma}^{*(2)}(E)+\left[\mathbf{-}\mathbf{u}+\mathbf{\Sigma}^{*(1)}(E)\right]\mathbf{G\mathnormal{^{0}(E)}}\left[\mathbf{-}\mathbf{u}+\mathbf{\Sigma}^{*(1)}(E)\right]\right\} +...,\label{sigmalambda}
\end{eqnarray}
which will be used to derive the perturbation energy functional.

\subsubsection{\emph{Perturbation energy up to the first order}}

We now consider Eq. \ref{sigmalambda} up to first order, $\mathbf{\Sigma}^{\lambda}(E)=\lambda\left[\mathbf{-}\mathbf{u}+\mathbf{\Sigma}^{*(1)}(E)\right]$.
Then, collecting the terms of order $\lambda^{0}$ in the integrand
of Eq. (\ref{eq:TotalE3}) gives

\begin{eqnarray}
E^{(1)} & = & E(0)+\frac{1}{2}\intop_{0}^{1}\frac{d\lambda}{\lambda}\int_{-\infty}^{+\infty}\frac{dE}{2\pi i}e^{iE\eta}\mathrm{Tr}\left\{ \lambda\left[\mathbf{-}2\mathbf{u}+\mathbf{\Sigma}^{*(1)}(E)\right]\mathbf{G}^{0}(E)\right\} \nonumber \\
 & = & \sum_{i}(h_{ii}+u_{ij})n_{i}+\left[-\sum_{i}u_{ii}n_{i}+\frac{1}{2}\sum_{k}\langle ik||ik\rangle n_{k}n_{i}\right]\nonumber \\
 & = & \sum_{i}h_{ii}n_{i}+\frac{1}{2}\sum_{ki}\langle ik||ik\rangle n_{k}n_{i}
\end{eqnarray}
which is just the Hartree-Fock energy functional for fractional systems.

\subsubsection{\emph{Perturbation energy up to the second order}}

Consider Eq. (\ref{sigmalambda}) up to second order and collect the
terms of order $\lambda^{1}$ in the integrand of Eq. (\ref{eq:TotalE3}),
then
\begin{eqnarray}
E^{(2)} & = & \frac{1}{2}\intop_{0}^{1}d\lambda\int_{-\infty}^{+\infty}\frac{dE}{2\pi i}e^{iE\eta}\mathrm{Tr}\{\mathbf{-}\mathbf{u}\left(\mathbf{G}^{0}(E)\left(\lambda\left[\mathbf{-}\mathbf{u}+\mathbf{\Sigma}^{*(1)}(E)\right]\right)\mathbf{G}^{0}(E)\right)\nonumber \\
 &  & +\frac{1}{\lambda}\lambda^{2}\left(\mathbf{\Sigma}^{*(2)}(E)+\left[\mathbf{-}\mathbf{u}+\mathbf{\Sigma}^{*(1)}(E)\right]\mathbf{G\mathnormal{^{0}(E)}}\left[\mathbf{-}\mathbf{u}+\mathbf{\Sigma}^{*(1)}(E)\right]\right)\mathbf{G}^{0}(E)\}\nonumber \\
 & = & \frac{1}{4}\int_{-\infty}^{+\infty}\frac{dE}{2\pi i}e^{iE\eta}\mathrm{Tr}\{\mathbf{\Sigma}^{*(2)}(E)\mathbf{G}^{0}(E)+\left[-2\mathbf{u}+\mathbf{\Sigma}^{*(1)}(E)\right]\mathbf{G\mathnormal{^{0}(E)}}\left[\mathbf{-}\mathbf{u}+\mathbf{\Sigma}^{*(1)}(E)\right]\mathbf{G}^{0}(E)\}
\end{eqnarray}
 Using Eq. (\ref{eq:2ndOrderE}),
\begin{eqnarray*}
 &  & \frac{1}{4}\int_{-\infty}^{+\infty}\frac{dE}{2\pi i}e^{iE\eta}\mathrm{Tr}\{\mathbf{\Sigma}^{*(2)}(E)\mathbf{G}^{0}(E)\}\\
 & = & \frac{1}{4}\sum_{ilmq}\langle iq||lm\rangle\langle lm||iq\rangle\frac{(1-n_{l})(1-n_{m})n_{q}n_{i}}{\varepsilon_{i}+\varepsilon_{q}-\varepsilon_{l}-\varepsilon_{m}}.
\end{eqnarray*}
 With HF orbitals where $\mathbf{-}\mathbf{u}+\mathbf{\Sigma}^{*(1)}(E)=0$,
the second order energy is given by
\begin{equation}
E^{(2)}=\frac{1}{4}\sum_{ilmq}\langle iq||lm\rangle\langle lm||iq\rangle\frac{(1-n_{l})(1-n_{m})n_{q}n_{i}}{\varepsilon_{i}+\varepsilon_{q}-\varepsilon_{l}-\varepsilon_{m}},\label{eq:MP2-frac}
\end{equation}
which is the fractional extension of the MP2 energy used previously
\cite{Cohen09786}. We attributed originally this fractional extension
to the finite-temperature extension of second-order perturbation theory
\cite{Blaizot86}, with the fractional occupations from finite temperature
excitations. However, the current derivation has been developed for
systems with fractional charges and spins at zero temperature.

Straightforwardly, we can obtain the derivatives with respect to the
occupation numbers of the MP2 second order energy
\begin{eqnarray}
\frac{\partial E^{(2)}}{\partial n_{i}} & = & \frac{1}{4}\sum_{lmq}\langle ij||lm\rangle\langle lm||ij\rangle2\frac{(1-n_{l})(1-n_{m})n_{j}n_{i}}{\varepsilon_{i}+\varepsilon_{j}-\varepsilon_{l}-\varepsilon_{m}}\nonumber \\
 &  & -\frac{1}{4}\sum_{jkm}\langle jk||im\rangle\langle lm||im\rangle2\frac{(1-n_{m})n_{j}n_{k}}{\varepsilon_{j}+\varepsilon_{k}-\varepsilon_{i}-\varepsilon_{m}}\nonumber \\
 & = & \frac{1}{2}\sum_{lmq}\langle iq||lm\rangle\langle lm||iq\rangle\left\{ \frac{(1-n_{l})(1-n_{m})n_{q}}{\varepsilon_{i}-\varepsilon_{l}-\varepsilon_{m}+\varepsilon_{q}}+\frac{n_{l}n_{m}(1-n_{q})}{\varepsilon_{i}-\varepsilon_{l}-\varepsilon_{m}+\varepsilon_{q}}\right\} \nonumber \\
 & = & {\Sigma}^{*(2)}(i,i,\varepsilon_{i}),
\end{eqnarray}
 where $\mathbf{-}\mathbf{u}+\mathbf{\Sigma}^{*(1)}(E)=0$ has again
been used. These derivatives have previously been used to calculate
the MP2 gap of some atoms and molecules \cite{Cohen09786}.

\section{Polarization Propagators and Random Phase Approximations}

Moving forward, we consider partial summation of perturbation theory
based on the random phase approximation. The particle-hole two-particle
Green's function is defined as
\begin{eqnarray}
G_{ph}(\mathbf{x}_{1},\mathbf{x}_{2},t_{1}-t_{2}) & = & -i\left\langle \Psi_{0}^{N}\left|T(\hat{\psi}^{\dagger}(1)\hat{\psi}(1)\hat{\psi}^{\dagger}(2)\hat{\psi}(2))\right|\Psi_{0}^{N}\right\rangle \nonumber \\
 & = & -i\left\langle \Psi_{0}^{N}\left|T(\hat{\rho}(1)\hat{\rho}(2))\right|\Psi_{0}^{N}\right\rangle \nonumber \\
 & = & -i\left\{ \theta(t_{1}-t_{2})\left\langle \Psi_{0}^{N}\left|\hat{\rho}(1)\hat{\rho}(2)\right|\Psi_{0}^{N}\right\rangle +\theta(t_{2}-t_{1})\left\langle \Psi_{0}^{N}\left|\hat{\rho}(2)\hat{\rho}(1)\right|\Psi_{0}^{N}\right\rangle \right\} \nonumber \\
 & = & -i\left\{ \theta(t_{1}-t_{2})\sum_{n}e^{i\left(E_{0}^{N}-E_{n}^{N}\right)(t_{1}-t_{2})}\left\langle \Psi_{0}^{N}\left|\hat{\rho}(\mathbf{x}_{1})\left|\Psi_{n}^{N}\right\rangle \left\langle \Psi_{n}^{N}\right|\hat{\rho}(\mathbf{x}_{2})\right|\Psi_{0}^{N}\right\rangle \right.\nonumber \\
 &  & \left.+\theta(t_{2}-t_{1})\sum_{n}e^{i\left(E_{0}^{N}-E_{n}^{N}\right)(t_{2}-t_{1})}\left\langle \Psi_{0}^{N}\left|\hat{\rho}(\mathbf{x}_{2})\left|\Psi_{n}^{N}\right\rangle \left\langle \Psi_{n}^{N}\right|\hat{\rho}(\mathbf{x}_{1})\right|\Psi_{0}^{N}\right\rangle \right\} .
\end{eqnarray}
The Lehmann representation in energy is given by
\begin{eqnarray}
G_{ph}(\mathbf{x}_{1},\mathbf{x}_{2},E) & = & \int_{-\infty}^{+\infty}d(t_{1}-t_{2})e^{iE(t_{1}-t_{2})}G_{ph}(\mathbf{x}_{1},\mathbf{x}_{2},t_{1}-t_{2})\nonumber \\
 & = & \sum_{n}\frac{\left\langle \Psi_{0}^{N}\left|\hat{\rho}(\mathbf{x}_{1})\left|\Psi_{n}^{N}\right\rangle \left\langle \Psi_{n}^{N}\right|\hat{\rho}(\mathbf{x}_{2})\right|\Psi_{0}^{N}\right\rangle }{E-(E_{n}^{N}-E_{0}^{N})+i\eta}-\sum_{n}\frac{\left\langle \Psi_{0}^{N}\left|\hat{\rho}(\mathbf{x}_{2})\left|\Psi_{n}^{N}\right\rangle \left\langle \Psi_{n}^{N}\right|\hat{\rho}(\mathbf{x}_{1})\right|\Psi_{0}^{N}\right\rangle }{E-(E_{0}^{N}-E_{n}^{N})-i\eta}.
\end{eqnarray}
Considering the density operator in terms of the field operators
\begin{eqnarray*}
\hat{\rho}(\mathbf{x}_{1}) & = & \hat{\psi}^{\dagger}(1)\hat{\psi}(1)\\
 & = & \sum_{j}\phi_{j}^{*}(\mathbf{x}_{1})a_{j}^{\dagger}(t_{1})\sum_{i}\phi_{i}(\mathbf{x}_{1})a_{i}(t_{1})
\end{eqnarray*}
 leads to the diagonal elements
\begin{equation}
G_{ph}(\mathbf{x}_{1},\mathbf{x}_{2};t_{1}-t_{2})=\sum_{ijkl}\phi_{j}^{*}(\mathbf{x}_{1})\phi_{i}(\mathbf{x}_{1})\phi_{k}^{*}(\mathbf{x}_{2})\phi_{l}(\mathbf{x}_{2})G_{ph}(i,j;k,l;t_{1}-t_{2}),
\end{equation}
and

\begin{equation}
G_{ph}(\mathbf{x}_{1},\mathbf{x}_{2};E)=\sum_{ijkl}\phi_{j}^{*}(\mathbf{x}_{1})\phi_{i}(\mathbf{x}_{1})\phi_{k}^{*}(\mathbf{x}_{2})\phi_{l}(\mathbf{x}_{2})G_{ph}(i,j;k,l;E),
\end{equation}
 where we have used the four-point Green's function
\begin{eqnarray}
G_{ph}(i,j;k,l;t_{1}-t_{2}) & = & -i\left\langle \Psi_{0}^{N}\left|T(a_{j}^{\dagger}(t_{1})a_{i}(t_{1})a_{k}^{\dagger}(t_{2})a_{l}(t_{2}))\right|\Psi_{0}^{N}\right\rangle \nonumber \\
 & = & -i\{\theta(t_{1}-t_{2})\sum_{n}e^{i\left(E_{0}^{N}-E_{n}^{N}\right)(t_{1}-t_{2})}\Big\langle\Psi_{0}^{N}\Big|a_{j}^{\dagger}a_{i}\Big|\Psi_{n}^{N}\Big\rangle\Big\langle\Psi_{n}^{N}\Big|a_{k}^{\dagger}a_{l}\Big|\Psi_{0}^{N}\Big\rangle\nonumber \\
 &  & +\theta(t_{2}-t_{1})\sum_{n}e^{i\left(E_{0}^{N}-E_{n}^{N}\right)(t_{2}-t_{1})}\Big\langle\Psi_{0}^{N}\Big|a_{k}^{\dagger}a_{l}\Big|\Psi_{n}^{N}\Big\rangle\Big\langle\Psi_{n}^{N}\Big|a_{j}^{\dagger}a_{i}\Big|\Psi_{0}^{N}\Big\rangle\}.
\end{eqnarray}
 Thus,
\begin{equation}
G_{ph}(i,j;k,l;E)=\sum_{n}\frac{\Big\langle\Psi_{0}^{N}\Big|a_{j}^{\dagger}a_{i}\Big|\Psi_{n}^{N}\Big\rangle\Big\langle\Psi_{n}^{N}\Big|a_{k}^{\dagger}a_{l}\Big|\Psi_{0}^{N}\Big\rangle}{E-(E_{n}^{N}-E_{0}^{N})+i\eta}-\sum_{n}\frac{\Big\langle\Psi_{0}^{N}\Big|a_{k}^{\dagger}a_{l}\Big|\Psi_{n}^{N}\Big\rangle\Big\langle\Psi_{n}^{N}\Big|a_{j}^{\dagger}a_{i}\Big|\Psi_{0}^{N}\Big\rangle}{E-(E_{0}^{N}-E_{n}^{N})-i\eta}.
\end{equation}

Similarly the polarization propagator in the spin and coordinate space
is defined as
\begin{eqnarray}
\Pi(\mathbf{x}_{1},\mathbf{x}_{2};t_{1}-t_{2}) & = & -i\left\{ \left\langle \Psi_{0}^{N}\left|T(\hat{\rho}(1)\hat{\rho}(2))\right|\Psi_{0}^{N}\right\rangle -\left\langle \Psi_{0}^{N}\left|\hat{\rho}(1)\right|\Psi_{0}^{N}\right\rangle \left\langle \Psi_{0}^{N}\left|\hat{\rho}(2)\right|\Psi_{0}^{N}\right\rangle \right\} \nonumber \\
 & = & -i\left\{ \theta(t_{1}-t_{2})\sum_{n\neq0}e^{i\left(E_{0}^{N}-E_{n}^{N}\right)(t_{1}-t_{2})}\left\langle \Psi_{0}^{N}\left|\hat{\rho}(\mathbf{x}_{1})\left|\Psi_{n}^{N}\right\rangle \left\langle \Psi_{n}^{N}\right|\hat{\rho}(\mathbf{x}_{2})\right|\Psi_{0}^{N}\right\rangle \right.\nonumber \\
 &  & \left.+\theta(t_{2}-t_{1})\sum_{n\neq0}e^{i\left(E_{0}^{N}-E_{n}^{N}\right)(t_{2}-t_{1})}\left\langle \Psi_{0}^{N}\left|\hat{\rho}(\mathbf{x}_{2})\left|\Psi_{n}^{N}\right\rangle \left\langle \Psi_{n}^{N}\right|\hat{\rho}(\mathbf{x}_{1})\right|\Psi_{0}^{N}\right\rangle \right\}
\end{eqnarray}
 or
\begin{equation}
\Pi(\mathbf{x}_{1},\mathbf{x}_{2};E)=\sum_{n\neq0}\frac{\left\langle \Psi_{0}^{N}\left|\hat{\rho}(\mathbf{x}_{1})\left|\Psi_{n}^{N}\right\rangle \left\langle \Psi_{n}^{N}\right|\hat{\rho}(\mathbf{x}_{2})\right|\Psi_{0}^{N}\right\rangle }{E-(E_{n}^{N}-E_{0}^{N})+i\eta}-\sum_{n\neq0}\frac{\left\langle \Psi_{0}^{N}\left|\hat{\rho}(\mathbf{x}_{2})\left|\Psi_{n}^{N}\right\rangle \left\langle \Psi_{n}^{N}\right|\hat{\rho}(\mathbf{x}_{1})\right|\Psi_{0}^{N}\right\rangle }{E-(E_{0}^{N}-E_{n}^{N})-i\eta}
\end{equation}
The four-point polarization propagators are
\begin{eqnarray}
\Pi(i,j;k,l;t_{1}-t_{2}) & = & -i\left\{ \Big\langle\Psi_{0}^{N}\Big|T(a_{j}^{\dagger}(t_{1})a_{i}(t_{1})a_{k}^{\dagger}(t_{2})a_{l}(t_{2}))\Big|\Psi_{0}^{N}\Big\rangle-\Big\langle\Psi_{0}^{N}\Big|a_{j}^{\dagger}a_{i}\Big|\Psi_{0}^{N}\Big\rangle\Big\langle\Psi_{0}^{N}\Big|a_{k}^{\dagger}a_{l}\Big|\Psi_{0}^{N}\Big\rangle\right\} \nonumber \\
 & = & -i\left\{ \theta(t_{1}-t_{2})\sum_{n}e^{i\left(E_{0}^{N}-E_{n}^{N}\right)(t_{1-}t_{2})}\Big\langle\Psi_{0}^{N}\Big|a_{j}^{\dagger}a_{i}\Big|\Psi_{n}^{N}\Big\rangle\Big\langle\Psi_{n}^{N}\Big|a_{k}^{\dagger}a_{l}\Big|\Psi_{0}^{N}\Big\rangle\right.\nonumber \\
 &  & \left.+\theta(t_{2}-t_{1})\sum_{n}e^{i\left(E_{0}^{N}-E_{n}^{N}\right)(t_{2}-t_{1})}\Big\langle\Psi_{0}^{N}\Big|a_{k}^{\dagger}a_{l}\Big|\Psi_{n}^{N}\Big\rangle\Big\langle\Psi_{n}^{N}\Big|a_{j}^{\dagger}a_{i}\Big|\Psi_{0}^{N}\Big\rangle\right\} \nonumber \\
 &  & -\Big\langle\Psi_{0}^{N}\Big|a_{j}^{\dagger}a_{i}\Big|\Psi_{0}^{N}\Big\rangle\Big\langle\Psi_{0}^{N}\Big|a_{k}^{\dagger}a_{l}\Big|\Psi_{0}^{N}\Big\rangle\nonumber \\
 & = & G_{ph}(j,i;k,l;t_{1}-t_{2})-\Big\langle\Psi_{0}^{N}\Big|a_{j}^{\dagger}a_{i}\Big|\Psi_{0}^{N}\Big\rangle\Big\langle\Psi_{0}^{N}\Big|a_{k}^{\dagger}a_{l}\Big|\Psi_{0}^{N}\Big\rangle,\label{eq:pi_t}
\end{eqnarray}
 or
\begin{eqnarray}
\Pi(i,j;k,l;E) & = & \sum_{n\neq0}\frac{\Big\langle\Psi_{0}^{N}\Big|a_{j}^{\dagger}a_{i}\Big|\Psi_{n}^{N}\Big\rangle\Big\langle\Psi_{n}^{N}\Big|a_{k}^{\dagger}a_{l}\Big|\Psi_{0}^{N}\Big\rangle}{E-(E_{n}^{N}-E_{0}^{N})+i\eta}-\sum_{n\neq0}\frac{\Big\langle\Psi_{0}^{N}\Big|a_{k}^{\dagger}a_{l}\Big|\Psi_{n}^{N}\Big\rangle\Big\langle\Psi_{n}^{N}\Big|a_{j}^{\dagger}a_{i}\Big|\Psi_{0}^{N}\Big\rangle}{E-(E_{0}^{N}-E_{n}^{N})-i\eta}\nonumber \\
 & = & G_{ph}(i,j;k,l;E)-2\pi\delta(E)\Big\langle\Psi_{0}^{N}\Big|a_{j}^{\dagger}a_{i}\Big|\Psi_{0}^{N}\Big\rangle\Big\langle\Psi_{0}^{N}\Big|a_{k}^{\dagger}a_{l}\Big|\Psi_{0}^{N}\Big\rangle.\label{eq:pi_E}
\end{eqnarray}
 In coordinate and spin representation $\mathbf{x}_{1}$, the diagonal
elements are
\begin{equation}
\Pi(\mathbf{x}_{1},\mathbf{x}_{2};t_{1}-t_{2})=\sum_{ijkl}\phi_{j}^{*}(\mathbf{x}_{1})\phi_{i}(\mathbf{x}_{1})\phi_{k}^{*}(\mathbf{x}_{2})\phi_{l}(\mathbf{x}_{2})\Pi(i,j;k,l;t_{1}-t_{2}),
\end{equation}
and

\begin{equation}
\Pi(\mathbf{x}_{1},\mathbf{x}_{2};E)=\sum_{ijkl}\phi_{j}^{*}(\mathbf{x}_{1})\phi_{i}(\mathbf{x}_{1})\phi_{k}^{*}(\mathbf{x}_{2})\phi_{l}(\mathbf{x}_{2})\Pi(i,j;k,l;E).
\end{equation}
The polarization propagator describes the dynamic density-density
fluctuation, which can be approximated in the random phase approximation
and leads to an approximate ground state correlation energy

\subsection{Non-interacting Systems}

The non-interacting propagator can be evaluated from Eqs. (\ref{eq:pi_t}
and \ref{eq:pi_E}), or it is given by \cite{Dickhoff05}
\begin{eqnarray}
\Pi^{0}(i,j;k,l;t_{1}-t_{2}) & = & -iG^{0}(i,k;t_{1}-t_{2})G^{0}(l,j;t_{2}-t_{1})\nonumber \\
 & = & -i(-i)\delta_{ik}e^{-i\varepsilon_{i}(t_{1}-t_{2})}\left\{ \theta(t_{1}-t_{2})\theta(i-F)-\theta(t_{2}-t_{1})\theta(F-i)\right\} \nonumber \\
 &  & (-i)\delta_{jl}e^{-i\varepsilon_{j}(t_{2}-t_{1})}\left\{ \theta(t_{2}-t_{1})\theta(j-F)-\theta(t_{1}-t_{2})\theta(F-j)\right\} \nonumber \\
 & = & -i\delta_{ik}\delta_{jl}e^{i(\varepsilon_{j}-\varepsilon_{i})(t_{1}-t_{2})}\left\{ \theta(t_{1}-t_{2})\theta(i-F)\theta(F-j)+\theta(t_{2}-t_{1})\theta(F-i)\theta(j-F)\right\} ,\label{eq:pi_0_t}
\end{eqnarray}
 and
\begin{eqnarray}
\Pi^{0}(i,j;k,l;E) & = & \int_{-\infty}^{+\infty}\frac{dE'}{2\pi i}G^{0}(i,k;E+E')G^{0}(l,j;E')\nonumber \\
 & = & \delta_{ik}\delta_{jl}\left\{ \frac{\theta(F-j)\theta(i-F)}{E-(\varepsilon_{i}-\varepsilon_{j})+i\eta}-\frac{\theta(F-i)\theta(j-F)}{E+(\varepsilon_{j}-\varepsilon_{i})-i\eta}\right\} \label{eq:pi_0_E}
\end{eqnarray}

\subsection{Non-interacting systems with fractional charge and fractional spin}

Using the key equations, Eqs. (\ref{eq:G_0_t_frac} and \ref{eq:G_0_E_frac}),
and following the same relationship as the integer cases in the previous
section, we obtain the propagator for fractional systems,
\begin{eqnarray}
\Pi^{0,N+\delta}(i,j;k,l;t_{1}-t_{2}) & = & -iG^{0,N+\delta}(i,k;t_{1}-t_{2})G^{0,N+\delta}(l,j;t_{2}-t_{1})\nonumber \\
 & = & -i\delta_{ik}\delta_{jl}e^{i(\varepsilon_{j}-\varepsilon_{i})(t_{1}-t_{2})}\left\{ \theta(t_{1}-t_{2})(1-n_{i})n_{j}+\theta(t_{2}-t_{1})n_{i}(1-n_{j})\right\} ,\label{eq:pi_0_t_frac}
\end{eqnarray}
and
\begin{eqnarray}
\Pi^{0,N+\delta}(i,j;k,l;E) & = & \int_{-\infty}^{+\infty}\frac{dE'}{2\pi i}G^{0,N+\delta}(i,k;E+E')G^{0,N+\delta}(l,j;E')\nonumber \\
 & = & \delta_{ik}\delta_{jl}\left\{ \frac{(1-n_{i})n_{j}}{E-(\varepsilon_{i}-\varepsilon_{j})+i\eta}-\frac{n_{i}(1-n_{j})}{E+(\varepsilon_{j}-\varepsilon_{i})-i\eta}\right\} .\label{eq:pi_0_E_frac}
\end{eqnarray}

This is another key equation, which raises several interesting points:
\begin{enumerate}
\item $\Pi^{0,N+\delta}$ is not the ensemble average of the $N$ and $N+\delta$
systems, whereas $G^{0,N+\delta}$ is.
\item In $\Pi^{0}(i,j;k,l;E)$, Eq. (\ref{eq:pi_0_E}), for integer systems
it is obtained that $\Pi^{0}(i,j;k,l;E)=(1-\delta_{ij})\Pi^{0}(i,j;E)$,
because of $\theta(F-j)\theta(i-F)$. However, for the $\Pi^{0,N+\delta}(i,j;k,l;E)$,
the term $i=j$ is non vanishing in general, because $(1-n_{i})n_{i}$
can be non-zero for fractionally occupied orbitals.
\item The index $i$ can be divided into three sets: particle ($p$) (or
unoccupied $u$), hole ($h$) (or occupied, $o$), and fractional
($f$). Note the somewhat confusing many-body terms of hole state
for $i<F$, and particle state for $F<i$. The particle and hole concept
is from the {}``vacuum'' of the Fermi sea, a particle is obtained
by adding it to the state above $F$ and hole is obtained for states
below $F$. In normal terms, the occupied states ($o$) are below
$F$, and the unoccupied states ($u$) are above $F$. When appropriate,
we will use $\{a,b,c,d,...\}$ to denote unoccupied (particle) and
fractional unoccupied states, $\{i,j,k,l,...\}$ to denote occupied
(hole) and fractional occupied states, and $\{m,n,o,p,...\}$ to denote
arbitrary states. In the concept of the {}``vacuum'' of the Fermi
sea, the fractional occupied states are both particle and hole states.
In the normal occupied/unoccupied concept, the fractional occupied
states are both occupied and unoccupied (fractionally).
\item $\Pi^{0,N+\delta}(i,j;k,l;E)$ is non-vanishing for the following
cases:

\begin{enumerate}
\item $(1-n_{i})n_{j}\neq0,$ when $j\leqslant F\leqslant i$; that is $j$
is a hole state ($j<F$) or a fractionally occupied state ($j=F$),
and $i$ is a particle ($F<i$) or a fractionally occupied state ($F=i$).
\item $n_{i}(1-n_{j})\neq0$, when $i\leqslant F\leqslant j$; that is $i$
is a hole state or a fractionally occupied state, $j$ is a particle
or a fractionally occupied state.
\end{enumerate}
\item $\Pi^{0,N+\delta}(i,j;k,l;E)$ is a diagonal matrix in the space of
$(ij\times kl)$ because of $\delta_{ik}\delta_{jl}$. Similarly to
Eq. (\ref{eq:Green_Matrix_frac}), we will use the unfolded matrix
representation of the propagator, whose dimension is $d_{ij}=2(n_{o}+n_{f})(n_{u}+n_{f})$,
where $n_{o}$ is the number of (fully) occupied orbitals, $n_{f}$
the number of fractionally occupied orbitals, and $n_{u}$ the number
of fully unoccupied orbitals. Its structure is given in Table 1.
\end{enumerate}
\begin{table}
\caption{Structure of $\Pi^{0,N+\delta}(i,j;k,l;E)$}

\begin{tabular}{c||c|c|c|c}
\hline
 & \multicolumn{4}{c}{$j$ index}\tabularnewline
\hline
$i$ index  & o (occupied)  & of ( occupied fractionally)  & uf ( unoccupied fractionally)  & u (unoccupied)\tabularnewline
\hline
\hline
o  & $0$  & $0$  & $-\frac{(1-n_{j})}{E+(\varepsilon_{j}-\varepsilon_{i})-i\eta}$  & $-\frac{1}{E+(\varepsilon_{j}-\varepsilon_{i})-i\eta}$\tabularnewline
\hline
of  & $0$  & $0$  & $-\frac{n_{i}(1-n_{j})}{E+(\varepsilon_{j}-\varepsilon_{i})-i\eta}$ & $-\frac{n_{i}}{E+(\varepsilon_{j}-\varepsilon_{i})-i\eta}$\tabularnewline
\hline
uf  & $\frac{(1-n_{i})}{E-(\varepsilon_{i}-\varepsilon_{j})+i\eta}$  & $\frac{(1-n_{i})n_{j}}{E-(\varepsilon_{i}-\varepsilon_{j})+i\eta}$ & $0$  & $0$\tabularnewline
\hline
u  & $\frac{1}{E-(\varepsilon_{i}-\varepsilon_{j})+i\eta}$  & $\frac{n_{j}}{E-(\varepsilon_{i}-\varepsilon_{j})+i\eta}$  & $0$  & $0$\tabularnewline
\hline
\end{tabular}
\end{table}

In the two-point spin coordinate representation, the fractional propagator
is given by
\begin{eqnarray}
\Pi^{0,N+\delta}(\mathbf{x}_{1},\mathbf{x}_{2};t_{1}-t_{2}) & = & \sum_{ijkl}\phi_{j}^{*}(\mathbf{x}_{1})\phi_{i}(\mathbf{x}_{1})\phi_{k}^{*}(\mathbf{x}_{2})\phi_{l}(\mathbf{x}_{2})\Pi^{0,N+\delta}(i,j;k,l;t_{1}-t_{2}),
\end{eqnarray}
 which in the energy form is
\begin{eqnarray*}
\Pi^{0,N+\delta}(\mathbf{x}_{1},\mathbf{x}_{2};E) & = & \sum_{ijkl}\phi_{j}^{*}(\mathbf{x}_{1})\phi_{i}(\mathbf{x}_{1})\phi_{k}^{*}(\mathbf{x}_{2})\phi_{l}(\mathbf{x}_{2})\Pi^{0,N+\delta}(i,j;k,l;E)\\
 & = & \sum_{ij}\phi_{j}^{*}(\mathbf{x}_{1})\phi_{i}(\mathbf{x}_{1})\phi_{i}^{*}(\mathbf{x}_{2})\phi_{j}(\mathbf{x}_{2})\left\{ \frac{(1-n_{i})n_{j}}{E-(\varepsilon_{i}-\varepsilon_{j})+i\eta}-\frac{n_{i}(1-n_{j})}{E+(\varepsilon_{j}-\varepsilon_{i})-i\eta}\right\} \\
 & = & \sum_{ij}\frac{\bar{\rho}_{i}(\mathbf{x}_{1},\mathbf{x}_{2})\rho_{j}(\mathbf{x}_{2},\mathbf{x}_{1})}{E-(\varepsilon_{i}-\varepsilon_{j})+i\eta}-\frac{\rho_{i}(\mathbf{x}_{1},\mathbf{x}_{2})\bar{\rho_{j}}(\mathbf{x}_{2},\mathbf{x}_{1})}{E+(\varepsilon_{j}-\varepsilon_{i})-i\eta},
\end{eqnarray*}
 where the occupied density matrix (hole) is
\begin{equation}
\rho_{i}(\mathbf{x}_{1},\mathbf{x}_{2})=n_{i}\phi_{i}(\mathbf{x}_{1})\phi_{i}^{*}(\mathbf{x}_{2})
\end{equation}
 and the unoccupied density matrix (particle) is
\begin{equation}
\bar{\rho}_{i}(\mathbf{x}_{1},\mathbf{x}_{2})=(1-n_{i})\phi_{i}(\mathbf{x}_{1})\phi_{i}^{*}(\mathbf{x}_{2})
\end{equation}
 If the orbitals are real, then
\begin{equation}
\rho_{i}(\mathbf{x}_{1},\mathbf{x}_{2})=\rho_{i}(\mathbf{x}_{2},\mathbf{x}_{1})
\end{equation}
 and the second part in $\Pi^{0,N+\delta}(\mathbf{x}_{1},\mathbf{x}_{2};E)$
can be rearranged as
\begin{eqnarray}
\sum_{ij}\frac{\rho_{i}(\mathbf{x}_{1},\mathbf{x}_{2})\bar{\rho_{j}}(\mathbf{x}_{2},\mathbf{x}_{1})}{E+(\varepsilon_{j}-\varepsilon_{i})-i\eta} & = & \sum_{ij}\frac{\rho_{j}(\mathbf{x}_{1},\mathbf{x}_{2})\bar{\rho_{i}}(\mathbf{x}_{2},\mathbf{x}_{1})}{E+(\varepsilon_{i}-\varepsilon_{j})-i\eta}\nonumber \\
 & = & \sum_{ij}\frac{\rho_{j}(\mathbf{x}_{2},\mathbf{x}_{1})\bar{\rho_{i}}(\mathbf{x}_{1},\mathbf{x}_{2})}{E+(\varepsilon_{i}-\varepsilon_{j})-i\eta}
\end{eqnarray}
 therefore
\begin{equation}
\Pi^{0,N+\delta}(\mathbf{x}_{1},\mathbf{x}_{2};E)=\sum_{ij}\bar{\rho}_{i}(\mathbf{x}_{1},\mathbf{x}_{2})\rho_{j}(\mathbf{x}_{2},\mathbf{x}_{1})\left(\frac{1}{E-(\varepsilon_{i}-\varepsilon_{j})+i\eta}-\frac{1}{E+(\varepsilon_{i}-\varepsilon_{j})-i\eta}\right)\label{pi0npd}
\end{equation}

\subsection{Ground State Energy from the Polarization Propagators}

A nice property of the polarization propagator is that it is connected
directly to ground state density matrix and hence the total energy.

\subsubsection{Potential Energy from \textmd{\textup{$\Pi(\mathbf{x},\mathbf{y};E)$}}}

Consider the interaction energies
\begin{eqnarray}
\left\langle \Psi_{0}^{N}\left|V\right|\Psi_{0}^{N}\right\rangle  & = & \left\langle \Psi_{0}^{N}(1)\left|V\right|\Psi_{0}^{N}(1)\right\rangle \nonumber \\
 & = & \intop_{0}^{1}d\lambda\left\langle \Psi_{0}^{N}(\lambda)\left|V\right|\Psi_{0}^{N}(\lambda)\right\rangle +\left\langle \Psi_{0}^{N}(0)\left|V\right|\Psi_{0}^{N}(0)\right\rangle ,
\end{eqnarray}
 and
\begin{eqnarray}
\left\langle \Psi_{0}^{N}(\lambda)\left|\lambda V\right|\Psi_{0}^{N}(\lambda)\right\rangle  & = & \frac{1}{2}\lambda\sum_{ijkl}\langle ij|kl\rangle\left\langle \Psi_{0}^{N}(\lambda)\left|a_{i}^{\dagger}(t)a_{j}^{\dagger}(t)a_{l}(t)a_{k}(t)\right|\Psi_{0}^{N}(\lambda)\right\rangle \nonumber \\
 & = & \frac{1}{2}\lambda\int\int d\mathbf{x}d\mathbf{y}v(\mathbf{x},\mathbf{y})\left\langle \Psi_{0}^{N}(\lambda)\left|\hat{\psi}^{\dagger}(\mathbf{x},t)\hat{\psi^{\dagger}}(\mathbf{y},t)\hat{\psi}(\mathbf{y},t)\hat{\psi}(\mathbf{x},t)\right|\Psi_{0}^{N}(\lambda)\right\rangle .
\end{eqnarray}

The diagonal part of the two particle reduced density matrix (2-rdm)
is
\begin{eqnarray}
 &  & 2\gamma_{2}(\mathbf{y},\mathbf{x})\nonumber \\
 & = & \left\langle \Psi_{0}^{N}\left|\hat{\psi}^{\dagger}(\mathbf{x})\hat{\psi^{\dagger}}(\mathbf{y})\hat{\psi}(\mathbf{y})\hat{\psi}(\mathbf{x})\right|\Psi_{0}^{N}\right\rangle \nonumber \\
 & = & \left\langle \Psi_{0}^{N}\left|-\hat{\psi}^{\dagger}(\mathbf{x})\hat{\psi^{\dagger}}(\mathbf{y})\hat{\psi}(\mathbf{x})\hat{\psi}(\mathbf{y})\right|\Psi_{0}^{N}\right\rangle \nonumber \\
 & = & \left\langle \Psi_{0}^{N}\left|-\hat{\psi}^{\dagger}(\mathbf{x})\left(\delta(\mathbf{x}-\mathbf{y})-\hat{\psi}(\mathbf{x})\hat{\psi^{\dagger}}(\mathbf{y})\right)\hat{\psi}(\mathbf{y})\right|\Psi_{0}^{N}\right\rangle \nonumber \\
 & = & -\delta(\mathbf{x}-\mathbf{y})\left\langle \Psi_{0}^{N}\left|\hat{\psi}^{\dagger}(\mathbf{x})\hat{\psi}(\mathbf{y})\right|\Psi_{0}^{N}\right\rangle \nonumber \\
 &  & +\left\langle \Psi_{0}^{N}\left|\hat{\psi}^{\dagger}(\mathbf{x})\left(\hat{\psi}(\mathbf{x})\hat{\psi^{\dagger}}(\mathbf{y})\right)\hat{\psi}(\mathbf{y})\right|\Psi_{0}^{N}\right\rangle \nonumber \\
 & = & -\delta(\mathbf{x}-\mathbf{y})\rho(\mathbf{x})+\left\langle \Psi_{0}^{N}\left|\hat{\rho}(\mathbf{x})\hat{\rho}(\mathbf{y})\right|\Psi_{0}^{N}\right\rangle \nonumber \\
 & = & \rho(\mathbf{x})\rho(\mathbf{y})-\delta(\mathbf{x}-\mathbf{y})\rho(\mathbf{x})+\left\langle \Psi_{0}^{N}\left|\left(\hat{\rho}(\mathbf{x})-\rho(\mathbf{x})\right)\left(\hat{\rho}(\mathbf{y})-\rho(\mathbf{y})\right)\right|\Psi_{0}^{N}\right\rangle \nonumber \\
 & = & \rho(\mathbf{x})\rho(\mathbf{y})-\delta(\mathbf{x}-\mathbf{y})\rho(\mathbf{x})+\sum_{n\neq0}\left\langle \Psi_{0}^{N}\left|\hat{\rho}(\mathbf{x})\right|\Psi_{n}^{N}\right\rangle \left\langle \Psi_{n}^{N}\left|\hat{\rho}(\mathbf{y})\right|\Psi_{0}^{N}\right\rangle \label{eq:2rdm_diag}
\end{eqnarray}

The last term in the previous equation can be rewritten as
\begin{eqnarray}
 &  & \sum_{n\neq0}\left\langle \Psi_{0}^{N}\left|\hat{\rho}(\mathbf{x})\right|\Psi_{n}^{N}\right\rangle \left\langle \Psi_{n}^{N}\left|\hat{\rho}(\mathbf{y})\right|\Psi_{0}^{N}\right\rangle \nonumber \\
 & = & \sum_{n\neq0}\int_{0}^{\infty}dE\frac{-1}{\pi}\textrm{Im}\int\frac{\left\langle \Psi_{0}^{N}\left|\hat{\rho}(\mathbf{x})\right|\Psi_{n}^{N}\right\rangle \left\langle \Psi_{n}^{N}\left|\hat{\rho}(\mathbf{y})\right|\Psi_{0}^{N}\right\rangle }{E-(E_{n}^{N}-E_{0}^{N})+i\eta}\nonumber \\
 & = & \sum_{n\neq0}\int_{0}^{\infty}dE\frac{-1}{\pi}\textrm{Im}\left[\int\frac{\left\langle \Psi_{0}^{N}\left|\hat{\rho}(\mathbf{x})\right|\Psi_{n}^{N}\right\rangle \left\langle \Psi_{n}^{N}\left|\hat{\rho}(\mathbf{y})\right|\Psi_{0}^{N}\right\rangle }{E-(E_{n}^{N}-E_{0}^{N})+i\eta}-\int\frac{\left\langle \Psi_{0}^{N}\left|\hat{\rho}(\mathbf{y})\right|\Psi_{n}^{N}\right\rangle \left\langle \Psi_{n}^{N}\left|\hat{\rho}(\mathbf{x})\right|\Psi_{0}^{N}\right\rangle }{E+(E_{n}^{N}-E_{0}^{N})-i\eta}\right]\nonumber \\
 & = & \sum_{n\neq0}\int_{-\infty}^{\infty}dE\frac{-1}{2\pi}\textrm{Im}\left[\frac{\left\langle \Psi_{0}^{N}\left|\hat{\rho}(\mathbf{x})\right|\Psi_{n}^{N}\right\rangle \left\langle \Psi_{n}^{N}\left|\hat{\rho}(\mathbf{y})\right|\Psi_{0}^{N}\right\rangle }{E-(E_{n}^{N}-E_{0}^{N})+i\eta}-\frac{\left\langle \Psi_{0}^{N}\left|\hat{\rho}(\mathbf{y})\right|\Psi_{n}^{N}\right\rangle \left\langle \Psi_{n}^{N}\left|\hat{\rho}(\mathbf{x})\right|\Psi_{0}^{N}\right\rangle }{E+(E_{n}^{N}-E_{0}^{N})-i\eta}\right]\nonumber \\
 & = & \frac{-1}{2\pi}\int_{-\infty}^{\infty}dE\textrm{Im}\Pi(\mathbf{x},\mathbf{y};E)\nonumber \\
 & = & \frac{-1}{2\pi i}\int_{-i\infty}^{i\infty}dEe^{\pm\eta}\Pi(\mathbf{x},\mathbf{y};E)\label{eq:Pi_integral}
\end{eqnarray}
 where we have used $\sum_{n\neq0}\left\langle \Psi_{0}^{N}\left|\hat{\rho}(\mathbf{x})\right|\Psi_{n}^{N}\right\rangle \left\langle \Psi_{n}^{N}\left|\hat{\rho}(\mathbf{y})\right|\Psi_{0}^{N}\right\rangle =\sum_{n\neq0}\left\langle \Psi_{0}^{N}\left|\hat{\rho}(\mathbf{y})\right|\Psi_{n}^{N}\right\rangle \left\langle \Psi_{n}^{N}\left|\hat{\rho}(\mathbf{x})\right|\Psi_{0}^{N}\right\rangle $
as the 2-rdm is symmetric, $\gamma_{2}(\mathbf{y},\mathbf{x})=\gamma_{2}(\mathbf{x},\mathbf{y})$

Now from the previous equations, we have
\begin{eqnarray}
2\gamma_{2}(\mathbf{y},\mathbf{x}) & = & \rho(\mathbf{x})\rho(\mathbf{y})-\delta(\mathbf{x}-\mathbf{y})\rho(\mathbf{x})-\frac{1}{2\pi}\int_{-\infty}^{\infty}dE\textrm{Im}\Pi(\mathbf{x},\mathbf{y};E)\nonumber \\
 & = & \rho(\mathbf{x})\rho(\mathbf{y})-\delta(\mathbf{x}-\mathbf{y})\rho(\mathbf{x})-\frac{1}{2\pi i}\int_{-i\infty}^{i\infty}dEe^{\pm i\eta}\Pi(\mathbf{x},\mathbf{y};E).\label{eq:Pi_integral2}
\end{eqnarray}
 Hence
\begin{eqnarray}
 &  & \left\langle \Psi_{0}^{N}\left|V\right|\Psi_{0}^{N}\right\rangle \nonumber \\
 & = & \int\int d\mathbf{x}d\mathbf{y}v(\mathbf{x},\mathbf{y})\gamma_{2}(\mathbf{y},\mathbf{x})\\
 & = & \frac{1}{2}\int\int d\mathbf{x}d\mathbf{y}v(\mathbf{x},\mathbf{y})\left[-\frac{1}{2\pi}\int_{-\infty}^{\infty}dE\textrm{Im}\Pi(\mathbf{x},\mathbf{y}_{2};E)+\rho(\mathbf{x})\rho(\mathbf{y})-\delta(\mathbf{x}-\mathbf{y})\rho(\mathbf{x})\right]\nonumber \\
 & = & \frac{1}{2}\int\int d\mathbf{x}d\mathbf{y}v(\mathbf{x},\mathbf{y})\left[-\frac{1}{2\pi i}\int_{-i\infty}^{i\infty}dEe^{\pm i\eta}\Pi(\mathbf{x},\mathbf{y}_{2};E)+\rho(\mathbf{x})\rho(\mathbf{y})-\delta(\mathbf{x}-\mathbf{y})\rho(\mathbf{x})\right].\label{eq:P_E_contour}
\end{eqnarray}

\subsubsection{Potential Energy from $\Pi(i,j;k,l;E)$}

Sometime it is desirable to work in the 4-point index space. We can
express the potential energy as

\begin{eqnarray}
\left\langle \Psi_{0}^{N}\left|V\right|\Psi_{0}^{N}\right\rangle  & = & \frac{1}{2}\sum_{ijkl}\langle ij|kl\rangle\left\langle \Psi_{0}^{N}\left|a_{i}^{\dagger}a_{j}^{\dagger}a_{l}a_{k}\right|\Psi_{0}^{N}\right\rangle \nonumber \\
 & = & \frac{1}{4}\sum_{ijkl}\langle ij||kl\rangle\left\langle \Psi_{0}^{N}\left|a_{i}^{\dagger}a_{j}^{\dagger}a_{l}a_{k}\right|\Psi_{0}^{N}\right\rangle \nonumber \\
 & = & \sum_{ijkl}\langle ij|kl\rangle\gamma(ij,kl)\nonumber \\
 & = & \sum_{ijkl}\langle ij|\bar{V}_{ph}|kl\rangle\gamma(il,jk)\nonumber \\
 & = & \frac{1}{2}\sum_{ijkl}\langle ij||kl\rangle\gamma(ij,kl)\nonumber \\
 & = & \frac{1}{2}\sum_{ijkl}\langle ij|V_{ph}|kl\rangle\gamma(il,jk)\label{eq:P_E_0}
\end{eqnarray}
where we use
\begin{eqnarray*}
\gamma(ij,kl) & = & \frac{1}{2}\left\langle \Psi_{0}^{N}\left|a_{i}^{\dagger}a_{j}^{\dagger}a_{l}a_{k}\right|\Psi_{0}^{N}\right\rangle
\end{eqnarray*}

\begin{equation}
\langle ij|kl\rangle=\int\int d\mathbf{x}_{1}d\mathbf{x}_{2}\phi_{i}^{*}(\mathbf{x}_{1})\phi_{j}^{*}(\mathbf{x}_{2})v(r_{12})\phi_{k}(\mathbf{x}_{1})\phi_{l}(\mathbf{x}_{2})
\end{equation}
and
\begin{eqnarray}
\langle ij||kl\rangle & = & \langle ij|kl\rangle-\langle ij|lk\rangle
\end{eqnarray}

\begin{eqnarray}
\langle ij|V_{ph}|kl\rangle & = & \langle il||jk\rangle\nonumber \\
 & = & \int\int d\mathbf{x}_{1}d\mathbf{x}_{2}\phi_{i}^{*}(\mathbf{x}_{1})\phi_{l}^{*}(\mathbf{x}_{2})v(r_{12})(1-P_{12})\phi_{j}(\mathbf{x}_{1})\phi_{k}(\mathbf{x}_{2})
\end{eqnarray}

\begin{eqnarray}
\langle ij|\bar{V}_{ph}|kl\rangle & = & \langle il|jk\rangle\nonumber \\
 & = & \int\int d\mathbf{x}_{1}d\mathbf{x}_{2}\phi_{i}^{*}(\mathbf{x}_{1})\phi_{l}^{*}(\mathbf{x}_{2})v(r_{12})\phi_{j}(\mathbf{x}_{1})\phi_{k}(\mathbf{x}_{2})
\end{eqnarray}

Using Eq. (\ref{eq:pi_E}), we have

\begin{eqnarray*}
\frac{-1}{\pi}\int_{-\infty}^{\infty}dE\textrm{Im}\Pi(i,j;k,l;E) & = & \delta_{ki}\rho_{jl}+\delta_{lj}\rho_{ki}-2\rho_{kl}\rho_{ji}+4\gamma_{jk,il},
\end{eqnarray*}
which is derived in details in section (\ref{sub:The-integration-of_PI})
in the Appendix. Thus

\begin{eqnarray*}
2\gamma_{jk,il} & = & \frac{-1}{2\pi}\int_{-\infty}^{\infty}dE\textrm{Im}\Pi(i,j;k,l;E)+\rho_{kl}\rho_{ji}-\frac{1}{2}\left(\delta_{ki}\rho_{jl}+\delta_{lj}\rho_{ki}\right)
\end{eqnarray*}
and for real orbitals, $\gamma_{jk,il}=\gamma_{il,jk}$ .

We now can express the potential energy as

\begin{eqnarray}
\left\langle \Psi_{0}^{N}\left|V\right|\Psi_{0}^{N}\right\rangle  & = & \sum_{ijkl}\langle ij|\bar{V}_{ph}|kl\rangle\gamma_{il,jk}\nonumber \\
 & = & \frac{1}{2}\sum_{ijkl}\langle ij|\bar{V}_{ph}|kl\rangle\left[\frac{-1}{2\pi}\int_{-\infty}^{\infty}dE\textrm{Im}\Pi(i,j;k,l;E)+\rho_{kl}\rho_{ji}-\frac{1}{2}\left(\delta_{ki}\rho_{jl}+\delta_{lj}\rho_{ki}\right)\right]\nonumber \\
 & = & \frac{-1}{4\pi}\int_{-\infty}^{\infty}dE\textrm{Im}\textrm{Tr }\mathbf{\bar{V}}_{ph}\Pi+\frac{1}{2}\int\int d\mathbf{x}d\mathbf{y}v(\mathbf{x},\mathbf{y})\left[\rho(\mathbf{x})\rho(\mathbf{y})-\delta(\mathbf{x}-\mathbf{y})\rho(\mathbf{x})\right]\label{eq:P_E_1}
\end{eqnarray}
Equivalently,
\begin{eqnarray}
\left\langle \Psi_{0}^{N}\left|V\right|\Psi_{0}^{N}\right\rangle  & = & \frac{1}{2}\sum_{ijkl}\langle ij|V_{ph}|kl\rangle\gamma_{il,jk}\nonumber \\
 & = & \frac{-1}{8\pi}\int_{-\infty}^{\infty}dE\textrm{Im}\textrm{Tr }\mathbf{V}_{ph}\Pi\nonumber \\
 &  & +\frac{1}{4}\int\int d\mathbf{x}_{1}d\mathbf{x}_{2}\phi_{i}^{*}(\mathbf{x}_{1})\phi_{l}^{*}(\mathbf{x}_{2})v(\mathbf{x}_{1},\mathbf{x}_{2})(1-P_{12})\phi_{j}(\mathbf{x}_{1})\phi_{k}(\mathbf{x}_{2})\left[\rho_{kl}\rho_{ji}-\frac{1}{2}\left(\delta_{ki}\rho_{jl}+\delta_{lj}\rho_{ki}\right)\right]\nonumber \\
 & = & \frac{-1}{8\pi}\int_{-\infty}^{\infty}dE\textrm{Im}\textrm{Tr }\mathbf{V}_{ph}\Pi\nonumber \\
 &  & +\frac{1}{4}\int\int d\mathbf{x}_{1}d\mathbf{x}_{2}v(\mathbf{x}_{1},\mathbf{x}_{2})\left(\rho(\mathbf{x}_{1})\rho(\mathbf{x}_{2})-\rho(\mathbf{x}_{2},\mathbf{x}_{1})\rho(\mathbf{x}_{1},\mathbf{x}_{2})\right)\nonumber \\
 &  & +\frac{1}{4}\int\int d\mathbf{x}_{1}d\mathbf{x}_{2}v(\mathbf{x}_{1},\mathbf{x}_{2})\left(\rho(\mathbf{x}_{1})\delta(\mathbf{x}_{2}-\mathbf{x}_{1})-\rho(\mathbf{x}_{1})\delta(0)\right).\label{eq:P_E_2}
\end{eqnarray}
See Section (\ref{sub:The-details-of_P_E}) in the Appendix for details
of the derivation.

\subsubsection{HF energy from $\Pi^{0,N+\delta}(\mathbf{x}_{1};\mathbf{x}_{2};E)$ }

Given the HF determinant wavefunction $\Phi_{0}^{N}$, the potential
energy in HF theory is given by
\begin{eqnarray}
E_{{\rm pot}}^{0} & = & \left\langle \Phi_{0}^{N}\left|V\right|\Phi_{0}^{N}\right\rangle \nonumber \\
 & = & \int\int d\mathbf{x}d\mathbf{y}v(\mathbf{x},\mathbf{y})\gamma_{2}(\mathbf{y},\mathbf{x})\nonumber \\
 & = & \frac{1}{2}\int\int d\mathbf{x}d\mathbf{y}v(\mathbf{x},\mathbf{y})\left[\rho(\mathbf{x})\rho(\mathbf{y})-\delta(\mathbf{x}-\mathbf{y})\rho(\mathbf{x})-\frac{1}{2\pi}\int_{-\infty}^{\infty}dE\textrm{Im}\Pi^{0}(\mathbf{x};\mathbf{y};E)\right]\label{epothf}
\end{eqnarray}
 To extend this expression to fractional charges and fractional spins,
simply use $\Pi^{0,N+\delta}(\mathbf{x}_{1};\mathbf{x}_{2};E)$ of
Eq. (\ref{pi0npd}), and
\begin{eqnarray}
-\frac{1}{2\pi}\int_{-\infty}^{\infty}dE\textrm{Im}\Pi^{0,N+\delta}(\mathbf{x}_{1};\mathbf{x}_{2};E) & = & -\frac{1}{2\pi}\int_{-\infty}^{\infty}dE\textrm{Im}\sum_{ij}\bar{\rho}_{i}(\mathbf{x}_{1},\mathbf{x}_{2})\rho_{j}(\mathbf{x}_{2},\mathbf{x}_{1})\nonumber \\
 &  & \left(\frac{1}{E-(\varepsilon_{i}-\varepsilon_{j})+i\eta}-\frac{1}{E+(\varepsilon_{i}-\varepsilon_{j})-i\eta}\right)\nonumber \\
 & = & \sum_{ij}\bar{\rho}_{i}(\mathbf{x}_{1},\mathbf{x}_{2})\rho_{j}(\mathbf{x}_{2},\mathbf{x}_{1})\nonumber \\
 & = & \bar{\rho}(\mathbf{x}_{1},\mathbf{x}_{2})\rho(\mathbf{x}_{2},\mathbf{x}_{1}).
\end{eqnarray}

Thus the last two terms of Eq. (\ref{epothf})
\begin{eqnarray*}
 &  & -\delta(\mathbf{\mathbf{x}}_{1}-\mathbf{x}_{2})\rho(\mathbf{x}_{1})-\frac{1}{2\pi}\int_{-\infty}^{\infty}dE\textrm{Im}\Pi^{0,N+\delta}(\mathbf{x}_{1};\mathbf{x}_{2};E)\\
 & = & -\delta(\mathbf{\mathbf{x}}_{1}-\mathbf{x}_{2})\rho(\mathbf{x}_{2},\mathbf{x}_{1})-\frac{1}{2\pi}\int_{-\infty}^{\infty}dE\textrm{Im}\Pi^{0,N+\delta}(\mathbf{x}_{1};\mathbf{x}_{2};E)\\
 & = & -\left(\bar{\rho}(\mathbf{x}_{1},\mathbf{x}_{2})+\rho(\mathbf{x}_{1},\mathbf{x}_{2})\right)\rho(\mathbf{x}_{2},\mathbf{x}_{1})+\bar{\rho}(\mathbf{x}_{1},\mathbf{x}_{2})\rho(\mathbf{x}_{2},\mathbf{x}_{1})\\
 & = & -\rho(\mathbf{x}_{1},\mathbf{x}_{2})\rho(\mathbf{x}_{2},\mathbf{x}_{1})
\end{eqnarray*}
 and we have
\begin{equation}
E_{{\rm pot}}^{0}=\frac{1}{2}\int\int d\mathbf{x}d\mathbf{y}v(\mathbf{x},\mathbf{y})\left[\rho(\mathbf{x})\rho(\mathbf{y})-\rho(\mathbf{x}_{1},\mathbf{x}_{2})\rho(\mathbf{x}_{2},\mathbf{x}_{1})\right],
\end{equation}
 which recovers the HF potential energy for fractional systems.

\subsubsection{Correlation energy }

We will use an adiabatic connection, which is more general than Eq.
(\ref{eq:AC_1}), to calculate the total energy.

\begin{equation}
\left\langle \Psi_{0}^{N}(\lambda)\left|H_{1}\right|\Psi_{0}^{N}(\lambda)\right\rangle =\int\int d\mathbf{x}d\mathbf{y}v(\mathbf{x},\mathbf{y})\gamma_{2}^{\lambda}(\mathbf{y},\mathbf{x})
\end{equation}
 with corresponding energy
\begin{equation}
E(1)=E(0)+\intop_{0}^{1}d\lambda\left\langle \Psi_{0}^{N}(\lambda)\left|H_{1}\right|\Psi_{0}^{N}(\lambda)\right\rangle
\end{equation}

We use a new expression for $H_{1}(\lambda)$, $H(\lambda)=H_{0}+H_{1}(\lambda)$,
such that$H(1)=H_{0}+H_{1}(1)$ is the physical $H$.

\[
E(1)=E(0)+\intop_{0}^{1}d\lambda\left\langle \Psi_{0}^{N}(\lambda)\left|\frac{\partial H_{1}}{\partial\lambda}\right|\Psi_{0}^{N}(\lambda)\right\rangle
\]
Let $H(\lambda)=H_{0}+H_{1}(\lambda)$, $H(0)=H_{0}=\sum_{ij}\left(h_{ij}+u(1)_{ij}\right)a_{i}^{\dagger}a_{j}$,
and $H_{1}(\lambda)=\lambda V-\sum_{ij}u_{ij}(\lambda)a_{i}^{\dagger}a_{j}$;
then $E(0)=\textrm{Tr}\rho^{0}(h+u(1))$, and

\begin{eqnarray*}
 &  & E(1)-E(0)\\
 & = & \intop_{0}^{1}d\lambda\left[\left\langle \Psi_{0}^{N}(\lambda)\left|V\right|\Psi_{0}^{N}(\lambda)\right\rangle -\left\langle \Psi_{0}^{N}(\lambda)\left|\sum_{ij}\frac{\partial u_{ij}(\lambda)}{\partial\lambda}a_{i}^{\dagger}a_{j}\right|\Psi_{0}^{N}(\lambda)\right\rangle \right]\\
 & = & -\intop_{0}^{1}d\lambda\textrm{Tr}\rho^{\lambda}\left(\frac{\partial u(\lambda)}{\partial\lambda}\right)+\frac{1}{2}\intop_{0}^{1}d\lambda\int\int d\mathbf{x}d\mathbf{y}v(\mathbf{x},\mathbf{y})\rho^{\lambda}(\mathbf{x})\rho^{\lambda}(\mathbf{y})\\
 &  & -\frac{1}{2}\intop_{0}^{1}d\lambda\int\int d\mathbf{x}d\mathbf{y}v(\mathbf{x},\mathbf{y})\left[\delta(\mathbf{x}-\mathbf{y})\rho^{\lambda}(\mathbf{x})+\frac{1}{2\pi}\int_{-\infty}^{\infty}dE\textrm{Im}\Pi^{\lambda}(\mathbf{x};\mathbf{y};E)\right]\\
 & = & -\intop_{0}^{1}d\lambda\textrm{Tr}\rho^{\lambda}\left(\frac{\partial u(\lambda)}{\partial\lambda}\right)+\frac{1}{2}\intop_{0}^{1}d\lambda\int\int d\mathbf{x}d\mathbf{y}v(\mathbf{x},\mathbf{y})\left(\rho^{\lambda}(\mathbf{x})\rho^{\lambda}(\mathbf{y})\right)\\
 &  & -\frac{1}{2}\intop_{0}^{1}d\lambda\int\int d\mathbf{x}d\mathbf{y}v(\mathbf{x},\mathbf{y})\left[\delta(\mathbf{x}-\mathbf{y})\left(\rho^{\lambda}(\mathbf{x})-\rho^{0}(\mathbf{x})\right)+\frac{1}{2\pi}\int_{-\infty}^{\infty}dE\textrm{Im}\left(\textrm{ }\Pi^{\lambda}(\mathbf{x};\mathbf{y};E)-\Pi^{0}(\mathbf{x};\mathbf{y};E)\right)\right]\\
 &  & -\frac{1}{2}\int\int d\mathbf{x}d\mathbf{y}v(\mathbf{x},\mathbf{y})\left[\rho^{0}(\mathbf{x}_{1},\mathbf{x}_{2})\rho^{0}(\mathbf{x}_{2},\mathbf{x}_{1})\right]
\end{eqnarray*}
where we used the result from the previous section on $\Pi^{0}(\mathbf{x};\mathbf{y};E)$
:
\begin{eqnarray*}
 &  & \frac{1}{2}\int\int d\mathbf{x}d\mathbf{y}v(\mathbf{x},\mathbf{y})\left[\delta(\mathbf{x}-\mathbf{y})\rho^{0}(\mathbf{x})-\frac{1}{2\pi}\int_{-\infty}^{\infty}dE\textrm{Im}\Pi^{0}(\mathbf{x};\mathbf{y};E)\right]\\
 & = & \frac{1}{2}\int\int d\mathbf{x}d\mathbf{y}v(\mathbf{x},\mathbf{y})\left[\rho^{0}(\mathbf{x}_{1},\mathbf{x}_{2})\rho^{0}(\mathbf{x}_{2},\mathbf{x}_{1})\right]
\end{eqnarray*}

Thus we have the general expression for calculating the ground state
energy from polarization propagator:

\begin{eqnarray}
 &  & E(1)\nonumber \\
 & = & \textrm{Tr}\rho^{0}\left(h+u(1)\right)-\intop_{0}^{1}d\lambda\textrm{Tr}\rho^{\lambda}\left(\frac{\partial u(\lambda)}{\partial\lambda}\right)\nonumber \\
 &  & +\frac{1}{2}\intop_{0}^{1}d\lambda\int\int d\mathbf{x}d\mathbf{y}v(\mathbf{x},\mathbf{y})\left(\rho^{\lambda}(\mathbf{x})\rho^{\lambda}(\mathbf{y})-\rho^{0}(\mathbf{x},\mathbf{y})\rho^{0}(\mathbf{y},\mathbf{x})\right)\nonumber \\
 &  & -\frac{1}{2}\intop_{0}^{1}d\lambda\int\int d\mathbf{x}d\mathbf{y}v(\mathbf{x},\mathbf{y})(\delta(\mathbf{x}-\mathbf{y})\left(\rho^{\lambda}(\mathbf{x})-\rho^{0}(\mathbf{x})\right)\nonumber \\
 &  & +\frac{1}{2\pi}\int_{-\infty}^{\infty}dE\textrm{Im}\textrm{ }\left(\Pi^{\lambda}(\mathbf{x};\mathbf{y};E)-\Pi^{0}(\mathbf{x};\mathbf{y};E)\right)).\label{eq:E1}
\end{eqnarray}
In general $u(1)=(v_{s}-v),u(0)=0,$ where $v_{s}$ is the non-interacting
reference potential. In between both limits we can have any $u(\lambda)$,
so that $\intop_{0}^{1}d\lambda\rho^{\lambda}\left(\frac{\partial u(\lambda)}{\partial\lambda}\right)=\rho^{1}u(1)-\intop_{0}^{1}u(\lambda)\frac{\partial\rho^{\lambda}}{\partial\lambda}d\lambda$
, and
\begin{eqnarray*}
 &  & \textrm{Tr}\rho^{0}\left(h+u(1)\right)-\intop_{0}^{1}d\lambda\textrm{Tr}\rho^{\lambda}\left(\frac{\partial u(\lambda)}{\partial\lambda}\right)\\
 & = & \textrm{Tr}\rho^{0}h+\textrm{Tr}\left(\rho^{0}-\rho^{1}\right)u(1)+\intop_{0}^{1}d\lambda\textrm{Tr}u(\lambda)\frac{\partial\rho^{\lambda}}{\partial\lambda}.
\end{eqnarray*}
We then have another general expression for calculating the ground
state energy from the polarization propagator
\begin{eqnarray}
 &  & E(1)\nonumber \\
 & = & \textrm{Tr}\rho^{0}h+\textrm{Tr}\left(\rho^{0}-\rho^{1}\right)u(1)+\intop_{0}^{1}d\lambda\textrm{Tr}u(\lambda)\frac{\partial\rho^{\lambda}}{\partial\lambda}\nonumber \\
 &  & +\frac{1}{2}\intop_{0}^{1}d\lambda\int\int d\mathbf{x}d\mathbf{y}v(\mathbf{x},\mathbf{y})\left(\rho^{\lambda}(\mathbf{x})\rho^{\lambda}(\mathbf{y})-\rho^{0}(\mathbf{x},\mathbf{y})\rho^{0}(\mathbf{y},\mathbf{x})\right)\nonumber \\
 &  & -\frac{1}{2}\intop_{0}^{1}d\lambda\int\int d\mathbf{x}d\mathbf{y}v(\mathbf{x},\mathbf{y})\left[\delta(\mathbf{x}-\mathbf{y})\left(\rho^{\lambda}(\mathbf{x})-\rho^{0}(\mathbf{x})\right)\right.\nonumber \\
 &  & \left.+\frac{1}{2\pi}\int_{-\infty}^{\infty}dE\textrm{Im}\textrm{ }\left(\Pi^{\lambda}(\mathbf{x};\mathbf{y};E)-\Pi^{0}(\mathbf{x};\mathbf{y};E)\right)\right].\label{eq:E1_2}
\end{eqnarray}
There are various ways to design adiabatic connections \cite{Harris741170,Gunnarsson764274,Langreth772884,Yang9810107},
that are now considered in this context.

\paragraph{Linear potential connection}

First, one can follow the linear path for the potential $u(\lambda)=\lambda u(1)=\lambda(v_{s}-v)$
(\cite{Harris741170}), then
\begin{eqnarray*}
 &  & \textrm{Tr}\rho^{0}\left(h+u(1)\right)-\intop_{0}^{1}d\lambda\textrm{Tr}\rho^{\lambda}\left(\frac{\partial u(\lambda)}{\partial\lambda}\right)\\
 & = & \textrm{Tr}\rho^{0}\left(h+u(1)\right)-\intop_{0}^{1}d\lambda\textrm{Tr}\rho^{\lambda}\left(u(1)\right)\\
 & = & \textrm{Tr}\rho^{0}h+\textrm{\ensuremath{\intop_{0}^{1}}d\ensuremath{\lambda}Tr}\left(\rho^{0}-\rho^{\lambda}\right)u(1)\\
 & = & \textrm{Tr}\rho^{0}(t+v)+\textrm{\ensuremath{\intop_{0}^{1}}d\ensuremath{\lambda}Tr}\left(\rho^{0}-\rho^{\lambda}\right)(v_{s}-v)\\
 & = & \textrm{Tr}\rho^{0}t+\intop_{0}^{1}d\lambda\left[\textrm{Tr\ensuremath{\rho^{\lambda}}\ensuremath{v}\ensuremath{+}Tr}\left(\rho^{0}-\rho^{\lambda}\right)v_{s}\right]
\end{eqnarray*}
and the ground state total energy, Eq. (\ref{eq:E1_2}), for the linear
potential path is
\begin{eqnarray}
 &  & E(1)\nonumber \\
 & = & \textrm{Tr}\rho^{0}t+\intop_{0}^{1}d\lambda\left[\textrm{Tr\ensuremath{\rho^{\lambda}}\ensuremath{v}\ensuremath{+}Tr}\left(\rho^{0}-\rho^{\lambda}\right)v_{s}\right]\nonumber \\
 &  & +\frac{1}{2}\intop_{0}^{1}d\lambda\int\int d\mathbf{x}d\mathbf{y}v(\mathbf{x},\mathbf{y})\left(\rho^{\lambda}(\mathbf{x})\rho^{\lambda}(\mathbf{y})-\rho^{0}(\mathbf{x},\mathbf{y})\rho^{0}(\mathbf{y},\mathbf{x})\right)\nonumber \\
 &  & -\frac{1}{2}\intop_{0}^{1}d\lambda\int\int d\mathbf{x}d\mathbf{y}v(\mathbf{x},\mathbf{y})\left[\delta(\mathbf{x}-\mathbf{y})\left(\rho^{\lambda}(\mathbf{x})-\rho^{0}(\mathbf{x})\right)\right.\nonumber \\
 &  & \left.+\frac{1}{2\pi}\int_{-\infty}^{\infty}dE\textrm{Im}\textrm{ }\left(\Pi^{\lambda}(\mathbf{x};\mathbf{y};E)-\Pi^{0}(\mathbf{x};\mathbf{y};E)\right)\right].\label{eq:E1_HJ}
\end{eqnarray}

Using HF energy expression

\[
E^{HF}=\textrm{Tr}\rho^{0}h+\frac{1}{2}\intop_{0}^{1}d\lambda\int\int d\mathbf{x}d\mathbf{y}v(\mathbf{x},\mathbf{y})\left(\rho^{0}(\mathbf{x})\rho^{0}(\mathbf{y})-\rho^{0}(\mathbf{x},\mathbf{y})\rho^{0}(\mathbf{y},\mathbf{x})\right),
\]
 the correlation energy $E_{c}=E(1)-E^{HF}$ is then

\begin{eqnarray}
E_{c} & = & -\intop_{0}^{1}d\lambda\textrm{Tr}\left(\rho^{\lambda}-\rho^{0}\right)u+\frac{1}{2}\intop_{0}^{1}d\lambda\int\int d\mathbf{x}d\mathbf{y}v(\mathbf{x},\mathbf{y})\left(\rho^{\lambda}(\mathbf{x})\rho^{\lambda}(\mathbf{y})-\rho^{0}(\mathbf{x})\rho^{0}(\mathbf{y})\right)\nonumber \\
 &  & -\frac{1}{2}\intop_{0}^{1}d\lambda\int\int d\mathbf{x}d\mathbf{y}v(\mathbf{x},\mathbf{y})\left[\delta(\mathbf{x}-\mathbf{y})\left(\rho^{\lambda}(\mathbf{x})-\rho^{0}(\mathbf{x})\right)\right.\nonumber \\
 &  & \left.+\frac{1}{2\pi}\int_{-\infty}^{\infty}dE\textrm{Im}\textrm{ }\left(\Pi^{\lambda}(\mathbf{x};\mathbf{y};E)-\Pi^{0}(\mathbf{x};\mathbf{y};E)\right)\right].\label{eq:Ec}
\end{eqnarray}

\paragraph{Constant density connection}

The second path is the constant-density adiabatic connection (\cite{Gunnarsson764274,Langreth772884}),
$\frac{\partial\rho^{\lambda}}{\partial\lambda}=0,\rho^{0}=\rho^{\mbox{\ensuremath{\lambda}}}=\rho^{\mbox{1}}$,
namely, the electron density along the adiabatic connection is kept
constant and is equal to the density of the physical system. Then,
the ground state total energy, Eq. (\ref{eq:E1_2}), becomes

\begin{eqnarray*}
E(1) & = & \textrm{Tr}\rho h+\frac{1}{2}\int\int d\mathbf{x}d\mathbf{y}v(\mathbf{x},\mathbf{y})\left(\rho(\mathbf{x})\rho(\mathbf{y})-\rho^{0}(\mathbf{x},\mathbf{y})\rho^{0}(\mathbf{y},\mathbf{x})\right)\\
 &  & -\frac{1}{2}\intop_{0}^{1}d\lambda\int\int d\mathbf{x}d\mathbf{y}v(\mathbf{x},\mathbf{y})\left[\frac{1}{2\pi}\int_{-\infty}^{\infty}dE\textrm{Im}\textrm{ }\left(\Pi^{\lambda}(\mathbf{x};\mathbf{y};E)-\Pi^{0}(\mathbf{x};\mathbf{y};E)\right)\right].
\end{eqnarray*}
Thus the correlation energy is

\begin{eqnarray}
E_{c} & = & -\frac{1}{2}\intop_{0}^{1}d\lambda\int\int d\mathbf{x}d\mathbf{y}v(\mathbf{x},\mathbf{y})\left[\frac{1}{2\pi}\int_{-\infty}^{\infty}dE\textrm{Im}\textrm{ }\left(\Pi^{\lambda}(\mathbf{x};\mathbf{y};E)-\Pi^{0}(\mathbf{x};\mathbf{y};E)\right)\right].\label{eq:E_C_1}
\end{eqnarray}
In four-point matrix form, we have
\begin{eqnarray}
E_{c} & = & -\frac{1}{2}\intop_{0}^{1}d\lambda\frac{1}{2\pi}\int_{-\infty}^{\infty}dE\textrm{Im}\sum_{ijkl}\textrm{ }\textrm{ }\left(\Pi^{\lambda}(i,j;k,l;E)-\Pi^{0}(i,j;k,l;E)\right)\langle kl|\bar{V}_{ph}|ij\rangle\nonumber \\
 & = & -\frac{1}{2}\intop_{0}^{1}d\lambda\frac{1}{2\pi}\int_{-\infty}^{\infty}dE\textrm{Im Tr\textrm{ }\ensuremath{\left(\mathbf{\Pi}^{\lambda}(E)-\mathbf{\Pi}^{0}(E)\right)\bar{\mathbf{V}}_{ph}}}\nonumber \\
 & = & -\frac{1}{2}\intop_{0}^{1}d\lambda\frac{1}{2\pi}\int_{-i\infty}^{i\infty}dEe^{\pm i\eta}\textrm{ Tr\textrm{ }}\left(\mathbf{\Pi}^{\lambda}(E)-\mathbf{\Pi}^{0}(E)\right)\bar{\mathbf{V}}_{ph}.\label{eq:E_C_2}
\end{eqnarray}
and
\begin{eqnarray}
E_{c} & = & -\frac{1}{4}\intop_{0}^{1}d\lambda\frac{1}{2\pi}\int_{-\infty}^{\infty}dE\textrm{Im}\sum_{ijkl}\textrm{ }\textrm{ }\left(\Pi^{\lambda}(i,j;k,l;E)-\Pi^{0}(i,j;k,l;E)\right)\langle kl|V_{ph}|ij\rangle\nonumber \\
 & = & -\frac{1}{4}\intop_{0}^{1}d\lambda\frac{1}{2\pi}\int_{-\infty}^{\infty}dE\textrm{Im Tr}\textrm{ }\left(\mathbf{\Pi}^{\lambda}(E)-\mathbf{\Pi}^{0}(E)\right)\mathbf{V}_{ph}\nonumber \\
 & = & -\frac{1}{4}\intop_{0}^{1}d\lambda\frac{1}{2\pi}\int_{-i\infty}^{i\infty}dEe^{\pm i\eta}\textrm{ Tr\textrm{ }}\left(\mathbf{\Pi}^{\lambda}(E)-\mathbf{\Pi}^{0}(E)\right)\mathbf{V}_{ph}.\label{eq:E_C_3}
\end{eqnarray}
where we use the anti-symmetric interaction matrix $\langle il||jk\rangle=\langle ij|V_{ph}|kl\rangle$
to calculate the correlation energy, as in Eq. (\ref{eq:P_E_0}).
We also replaced the integration along the real axis by integration
along the imaginary axis, as they are shown equivalently in Eq. (\ref{eq:P_E_contour}).

Eq. (\ref{eq:E_C_1}) is convenient for RPA calculations without exchange
interaction in two-point real space representation, while Eq. (\ref{eq:E_C_2})
for RPA calculations without exchange interaction, and Eq. (\ref{eq:E_C_3})
for RPA calculations with exchange interaction.

\subsection{Random Phase Approximations}

\subsubsection{RPA Equations}

The random phase approximation (RPA) \cite{Negele88} can be written
as the equation for the polarization propagator,
\begin{eqnarray}
\Pi^{{\rm RPA}}(i,j;k,l;E) & = & \Pi^{0}(i,j;k,l;E)+\sum_{pqrs}\Pi^{0}(i,j;p,q;E)\langle pq|U_{ph}|rs\rangle\Pi^{{\rm RPA}}(r,s;k,l;E)\label{eq:RPA}
\end{eqnarray}
or in condensed notation
\begin{equation}
\mathbf{\Pi}^{{\rm RPA}}=\mathbf{\Pi}^{0}+\mathbf{\Pi}^{0}\mathbf{U}_{ph}\mathbf{\Pi}^{{\rm RPA}}\label{eq:RPAmatrix}
\end{equation}
where $\langle ij|U_{ph}|kl\rangle$ can be a general energy-independent
interaction. $\langle ij|U_{ph}|kl\rangle=\langle ij|\bar{V}_{ph}|kl\rangle$
for RPA with no exchange interaction, and $\langle ij|U_{ph}|kl\rangle=\langle ij|V_{ph}|kl\rangle$
for RPAE, the random phase approximation with exchange interaction.

Assume that $\Pi^{{\rm RPA}}$ has a Lehmann representation as that
of the exact one in Eq. (\ref{eq:pi_E}) \cite{Negele88} ,
\begin{eqnarray}
\Pi^{{\rm RPA}}(i,j;k,l;E) & = & \sum_{n\neq0}\frac{\Big\langle\Psi_{0}^{{\rm RPA}}\Big|a_{j}^{\dagger}a_{i}\Big|\Psi_{n}^{{\rm RPA}}\Big\rangle\Big\langle\Psi_{n}^{{\rm RPA}}\Big|a_{k}^{\dagger}a_{l}\Big|\Psi_{0}^{\mathrm{RPA}}\Big\rangle}{E-(E_{n}^{N}-E_{0}^{N})+i\eta}\nonumber \\
 &  & -\sum_{n\neq0}\frac{\Big\langle\Psi_{0}^{{\rm RPA}}\Big|a_{k}^{\dagger}a_{l}\Big|\Psi_{n}^{{\rm RPA}}\Big\rangle\Big\langle\Psi_{n}^{{\rm RPA}}\Big|a_{j}^{\dagger}a_{i}\Big|\Psi_{0}^{{\rm RPA}}\Big\rangle}{E-(E_{0}^{N}-E_{n}^{N})-i\eta}.
\end{eqnarray}
 Within RPA, define the excitation energy
\begin{equation}
\varepsilon_{n}^{\pi}=E_{n}^{N}-E_{0}^{N},
\end{equation}
 the column matrix $\mathbf{X}^{n}$ with dimension $(n_{u}+n_{f})(n_{o}+n_{f})$,
\begin{eqnarray}
X_{bi}^{n} & = & \left\langle \Psi_{n}^{{\rm RPA}}\right|a_{b}^{\dagger}a_{i}\left|\Psi_{0}^{{\rm RPA}}\right\rangle ^{*}\nonumber \\
 & = & \left\langle \Psi_{0}^{{\rm RPA}}\right|a_{i}^{\dagger}a_{b}\left|\Psi_{n}^{{\rm RPA}}\right\rangle
\end{eqnarray}
 and the column matrix $\mathbf{Y}^{n}$ with dimension $(n_{o}+n_{f})(n_{u}+n_{f})$,
\begin{eqnarray}
Y_{bi}^{n} & = & \left\langle \Psi_{n}^{{\rm RPA}}\right|a_{i}^{\dagger}a_{b}\left|\Psi_{0}^{{\rm RPA}}\right\rangle ,
\end{eqnarray}
then we can write \cite{Negele88}
\begin{equation}
\mathbf{\Pi}^{{\rm RPA}}=\sum_{n\neq0}\frac{\left(\begin{array}{c}
\mathbf{X}^{n}\\
\mathbf{Y}^{n}
\end{array}\right)(\mathbf{X}^{n\dagger}\mathbf{Y}^{n\dagger})}{E-\varepsilon_{n}^{\pi}+i\eta}-\sum_{n\neq0}\frac{\left(\left(\begin{array}{c}
\mathbf{Y}^{n}\\
\mathbf{X}^{n}
\end{array}\right)(\mathbf{Y}^{n\dagger}\mathbf{X}^{n\dagger})\right)^{*}}{E+\varepsilon_{n}^{\pi}-i\eta}.\label{eq:Pi_RPA_analytical}
\end{equation}

It is not necessary to solve the RPA equation, Eq. (\ref{eq:RPAmatrix}),
directly for every energy, because we can use the analytical structure
of $\mathbf{\Pi}(E)$ from Eq.(\ref{eq:Pi_RPA_analytical}). We multiply
Eq. (\ref{eq:RPAmatrix}) from the left by $\left(\mathbf{\Pi}^{0}\right)^{-1}$
,

\[
\left(\mathbf{\Pi}^{0}\right)^{-1}\mathbf{\mathbf{\Pi}^{{\rm RPA}}}=\mathbf{I}+\mathbf{U}_{ph}\mathbf{\mathbf{\Pi}^{{\rm RPA}},}
\]
thus,
\[
\left(\left(\mathbf{\Pi}^{0}\right)^{-1}-\mathbf{U}_{ph}\right)\mathbf{\mathbf{\mathbf{\Pi}^{{\rm RPA}}}}=\mathbf{I.}
\]
 We only need to solve the equations at the singularity of $\mathbf{\mathbf{\mathbf{\Pi}^{{\rm RPA}}}}$.
Thus
\[
\lim_{E\rightarrow\varepsilon_{n}^{\pi}-i\eta}\left(E-\varepsilon_{n}^{\pi}+i\eta\right)\left(\left(\mathbf{\Pi}^{0}\right)^{-1}-\mathbf{U}_{ph}\right)\mathbf{\mathbf{\mathbf{\Pi}^{{\rm RPA}}}}=0.
\]

Take the limit
\begin{equation}
\lim_{E\rightarrow\varepsilon_{n}^{\pi}-i\eta}\left(E-\varepsilon_{n}^{\pi}+i\eta\right)\mathbf{\Pi}^{{\rm RPA}}=\left(\begin{array}{c}
\mathbf{X}^{n}\\
\mathbf{Y}^{n}
\end{array}\right)(\mathbf{X}^{n\dagger}\mathbf{Y}^{n\dagger}),\label{limit0}
\end{equation}
\begin{equation}
\left(\left(\mathbf{\Pi}^{0}(\varepsilon_{n}^{\pi}-i\eta)\right)^{-1}-\mathbf{U}_{ph}\right)\left(\begin{array}{c}
\mathbf{X}^{n}\\
\mathbf{Y}^{n}
\end{array}\right)=0.\label{eq:RPAEqinXY}
\end{equation}
 Thus, Eq. (\ref{eq:RPAEqinXY}) is valid for any $\varepsilon_{n}^{\pi}$,
including the $\varepsilon_{n}^{\pi}=0$ when the ground state is
degenerate for both the many-electron and the non-interacting reference
system.

Eq. (\ref{eq:RPAEqinXY}) can be rewritten in terms of two blocks,
corresponding to the $\mathbf{X}^{n}$ and $\mathbf{Y}^{n}$ components:
\begin{eqnarray}
\left(\Pi^{0}(a,i;a,i;\varepsilon_{n}^{\pi}-i\eta)\right)^{-1}X_{ai}^{n} & = & \sum_{bj}\left(\langle ai|U_{ph}|bj\rangle X_{bj}^{n}+\langle ai|U_{ph}|jb\rangle Y_{bj}^{n}\right)\label{eq:RPA_de_1}
\end{eqnarray}
 and
\begin{equation}
\left(\Pi^{0}(i,a;i,a;\varepsilon_{n}^{\pi}-i\eta)\right)^{-1}Y_{ai}^{n}=\sum_{bj}\left(\langle ia|U_{ph}|bj\rangle X_{bj}^{n}+\langle ia|U_{ph}|jb\rangle Y_{bj}^{n}\right)\label{eq:RPA_de_2}
\end{equation}

Based on Table 1,$\left(\Pi^{0}(a,i;a,i;\varepsilon_{n}^{\pi}-i\eta)\right)^{-1}=\frac{\varepsilon_{n}^{\pi}-(\varepsilon_{a}-\varepsilon_{i})}{(1-n_{a})n_{i}}$,
then Eq. (\ref{eq:RPA_de_1}) becomes
\begin{eqnarray}
\frac{\varepsilon_{n}^{\pi}-(\varepsilon_{a}-\varepsilon_{i})}{(1-n_{a})n_{i}}X_{ai}^{n} & = & \frac{1}{\sqrt{(1-n_{a})n_{i}}}\sum_{bj}\left\{ \sqrt{(1-n_{a})n_{i}}\langle ai|U_{ph}|bj\rangle\sqrt{(1-n_{b})n_{j}}\frac{1}{\sqrt{(1-n_{b})n_{j}}}X_{bj}^{n}\right.\nonumber \\
 &  & \left.+\sqrt{(1-n_{a})n_{i}}\langle ai|U_{ph}|jb\rangle\sqrt{(1-n_{b})n_{j}}\frac{1}{\sqrt{(1-n_{b})n_{j}}}Y_{bj}^{n}\right\} .\label{xnaA}
\end{eqnarray}

Define the following fractional-transformed quantities
\begin{eqnarray}
\sqrt{(1-n_{a})n_{i}}\langle ai|U_{ph}|bj\rangle\sqrt{(1-n_{b})n_{j}} & = & \langle ai|U_{ph}|bj\rangle_{f},\\
\sqrt{(1-n_{a})n_{A}}\langle ai|U_{ph}|jb\rangle\sqrt{(1-n_{b})n_{j}} & = & \langle ai|U_{ph}|jb\rangle_{f},\\
\frac{1}{\sqrt{(1-n_{a})n_{i}}}X_{ai}^{n} & = & \widetilde{X}_{ai}^{n},\\
\frac{1}{\sqrt{(1-n_{b})n_{j}}}Y_{bj}^{n} & = & \widetilde{Y}_{bj}^{n}.
\end{eqnarray}
 Then Eq. (\ref{xnaA}) can be written as
\begin{equation}
\left(\varepsilon_{n}^{\pi}-(\varepsilon_{a}-\varepsilon_{i})\right)\widetilde{X}_{ai}^{n}=\sum_{bB}\left\{ \langle ai|U_{ph}|bj\rangle_{f}\widetilde{X}_{bj}^{n}+\langle ai|U_{ph}|jb\rangle_{f}\widetilde{Y}_{bj}^{n}\right\} .
\end{equation}

Similarly, using $\left(\Pi^{0}(i,a;i,a;\varepsilon_{n}^{\pi}-i\eta)\right)^{-1}=-\frac{\varepsilon_{n}^{\pi}+(\varepsilon_{a}-\varepsilon_{i})}{n_{i}(1-n_{a})}$
from Table 1, we can convert Eq. (\ref{eq:RPA_de_2}) into
\begin{equation}
-\left(\varepsilon_{n}^{\pi}+(\varepsilon_{a}-\varepsilon_{i})\right)\widetilde{Y}_{ai}^{n}=\sum_{bj}\{\langle ia|U_{ph}|bj\rangle_{f}\widetilde{X}_{bj}^{n}+\langle ia|U_{ph}|jb\rangle_{f}\widetilde{Y}_{bj}^{n}\}.
\end{equation}
 In matrix notation
\begin{equation}
\left(\begin{array}{cc}
\mathbf{A} & \mathbf{B}\\
-\mathbf{B^{*}} & -\mathbf{A^{*}}
\end{array}\right)\left(\begin{array}{c}
\mathbf{\widetilde{X}}^{n}\\
\widetilde{\mathbf{Y}}^{n}
\end{array}\right)=\varepsilon_{n}^{\pi}\left(\begin{array}{c}
\mathbf{\widetilde{X}}^{n}\\
\widetilde{\mathbf{Y}}^{n}
\end{array}\right)\label{eq:RPA_eig_frac}
\end{equation}
where the $A$ and $B$ matrices including fractional occupations
are
\begin{eqnarray}
A_{ia,jb} & = & (\varepsilon_{a}-\varepsilon_{i})\delta_{ab}\delta_{ij}+\langle ai|U_{ph}|bj\rangle\sqrt{(1-n_{a})n_{i}(1-n_{b})n_{j}},\label{eq:A_frac}
\end{eqnarray}
\begin{eqnarray}
B_{ia,jb} & = & \langle ab||ij\rangle\sqrt{(1-n_{a})n_{i}(1-n_{b})n_{j}}.\label{eq:B_frac}
\end{eqnarray}
This is the RPA equations for systems with fractional charges and
fractional spins, which were developed in our previous work \cite{Mori-Sanchez09,Mori-Sanchez1242507},
based on just using the occupation-scaled orbitals with Eqs. (\ref{eq:scaled_u}\ref{eq:scaled_o})
and extending the dimension of the matrices to include the fractional
orbitals in both the occupied and unoccupied states. The static limiting
case of $E=0$ has also been developed \cite{Peng130} in conjunction
with extending the analytical evaluation of Fukui functions \cite{Yang12144110}
and local conditions for the fractional charge and fractional spins.
Here we have given the full derivation starting from the basic formula
of the single-particle Green's function for fractional systems, Eq.(\ref{eq:G_0_E_frac}).
Our derivation also gives clear meaning for the eigenvectors $\mathbf{\widetilde{X}}^{n}$
and $\widetilde{\mathbf{Y}}^{n}$ for fractional systems.

\subsubsection{RPA Energy Expressions}

We now focus on the correlation energy. The RPA equation, Eq. (\ref{eq:RPAmatrix}),
for $\Pi^{\mathrm{RPA},\lambda}$ at the coupling constant $\lambda$
can be solved as\\
\[
\Pi^{\mathrm{RPA},\lambda}=\frac{1}{\left((\mathbf{\Pi^{0}})^{-1}-\lambda\mathbf{U}_{ph}\right)}.
\]
Then, the RPA approximation to correlation energy, based on Eq. (\ref{eq:E_C_2}),
is
\begin{eqnarray}
E_{c} & = & -\frac{1}{4\pi}\int_{-\infty}^{\infty}dE\textrm{Im}\intop_{0}^{1}d\lambda\textrm{Tr}\left[\frac{1}{\left((\mathbf{\Pi^{0}}(E))^{-1}-\lambda\mathbf{U}_{ph}\right)}-\mathbf{\Pi}^{0}(E)\right]\bar{\mathbf{V}}_{ph}\nonumber \\
 & = & -\frac{1}{4\pi}\int_{-\infty}^{\infty}dE\textrm{Im}\intop_{0}^{1}d\lambda\textrm{Tr}\left[\frac{1}{\left(1-\lambda\mathbf{\Pi}^{0}(E)\mathbf{U}_{ph}\right)}\mathbf{\Pi}^{0}(E)-\mathbf{\Pi}^{0}(E)\right]\bar{\mathbf{V}}_{ph}\nonumber \\
 &  & =-\frac{1}{4\pi}\int_{-\infty}^{\infty}dE\textrm{Im}\intop_{0}^{1}d\lambda\textrm{Tr}\left[1+\lambda\mathbf{\Pi}^{0}(E)\mathbf{U}_{ph}+\lambda^{2}\mathbf{\Pi}^{0}(E)\mathbf{U}_{ph}\mathbf{\Pi}^{0}(E)\mathbf{U}_{ph}+...-1\right]\mathbf{\Pi}^{0}(E)\bar{\mathbf{V}}_{ph}\nonumber \\
 &  & =-\frac{1}{4\pi}\int_{-\infty}^{\infty}dE\textrm{Im}\textrm{Tr}\left[\frac{1}{2}\mathbf{\Pi}^{0}(E)\mathbf{U}_{ph}+\frac{1}{3}\mathbf{\Pi}^{0}(E)\mathbf{U}_{ph}\mathbf{\Pi}^{0}(E)\mathbf{U}_{ph}...\right]\mathbf{\Pi}^{0}(E)\bar{\mathbf{V}}_{ph}\nonumber \\
 &  & =-\frac{1}{4\pi}\int_{-\infty}^{\infty}dE\textrm{Im}\textrm{Tr}\left[\frac{1}{2}\left(\mathbf{\Pi}^{0}(E)\mathbf{U}_{ph}\right)^{2}+\frac{1}{3}\left(\mathbf{\Pi}^{0}(E)\mathbf{U}_{ph}\right)^{3}+...\right]\left(\mathbf{U}_{ph}\right)^{-1}\bar{\mathbf{V}}_{ph}\nonumber \\
 &  & =\frac{1}{4\pi}\int_{-\infty}^{\infty}dE\textrm{Im}\textrm{Tr}\left[\ln(\mathbf{1}-\mathbf{\Pi}^{0}(E)\mathbf{U}_{ph})+\mathbf{\Pi}^{0}(E)\mathbf{U}_{ph}\right]\left(\mathbf{U}_{ph}\right)^{-1}\bar{\mathbf{V}}_{ph}.\label{eq:RPAcorrelation2}
\end{eqnarray}
Note that if we use $\langle jk||il\rangle=\langle kl|V_{ph}|ij\rangle$instead
of $\langle jk|il\rangle=\langle kl|\bar{V}_{ph}|ij\rangle$, there
is an additional factor of $\frac{1}{2}$ as in Eq. (\ref{eq:E_C_3}),
and therefore

\begin{eqnarray}
E_{c}^{\mathrm{KS-DFT}} & = & \frac{1}{8\pi}\int_{-\infty}^{\infty}dE\textrm{Im}\textrm{Tr}\left[\ln(\mathbf{1}-\mathbf{\Pi}^{0}(E)\mathbf{U}_{ph})+\mathbf{\Pi}^{0}(E)\mathbf{U}_{ph}\right]\left(\mathbf{U}_{ph}\right)^{-1}\mathbf{V}_{ph}.\label{eq:RPAcorrelation3}
\end{eqnarray}
We can see the convenience of using Eq. (\ref{eq:RPAcorrelation2})
for conventional RPA without exchange interaction where $\mathbf{U}_{ph}=\bar{\mathbf{V}}_{ph}$
and Eq. (\ref{eq:RPAcorrelation3}) for RPAE with exchange where $\mathbf{U}_{ph}=\mathbf{V}_{ph}$.
In Section (\ref{sub:RPA-correlation-energy}) of the Appendix, we
show that
\begin{eqnarray}
 &  & \frac{1}{4\pi}\int_{-\infty}^{\infty}dE\textrm{Im}\textrm{Tr}\left[\ln(\mathbf{1}-\mathbf{\Pi}^{0}(E)\mathbf{U}_{ph})+\mathbf{\Pi}^{0}(E)\mathbf{U}_{ph}\right]\nonumber \\
 & = & \frac{1}{2}\sum_{n>0}\varepsilon_{n}^{\pi}-\frac{1}{2}\sum_{a\in\mathrm{particle},i\in\mathrm{hole}}\left(\varepsilon_{a}-\varepsilon_{i}\right)-\frac{1}{2}\sum_{ij}\left\langle ij\right|U_{ph}\left|ij\right\rangle _{f}\nonumber \\
 & = & \frac{1}{2}\sum_{n>0}\varepsilon_{n}^{\pi}-\frac{1}{2}\textrm{Tr}\mathbf{A}\label{eq:E_integ}
\end{eqnarray}

Thus for RPA, $\mathbf{U}_{ph}=\bar{\mathbf{V}}_{ph}$ , without exchange
\begin{equation}
E_{c}^{{\rm RPA}}=\frac{1}{2}\sum_{n>0}\varepsilon_{n}^{\pi}-\frac{1}{2}\textrm{Tr}\mathbf{A}.\label{eq:RPAcorrelation}
\end{equation}
For RPAE, $\mathbf{U}_{ph}=\mathbf{V}_{ph}$, with exchange,
\begin{equation}
E_{c}^{{\rm RPAE}}=\frac{1}{4}\sum_{n>0}\varepsilon_{n}^{\pi}-\frac{1}{4}\textrm{Tr}\mathbf{A}\label{eq:RPAEcorrelation}
\end{equation}
These are the desired results rigorously derived from the ensemble
of the basic variable, which have been used previously in showing
the large localization error in the RPA energy \cite{Mori-Sanchez09,Mori-Sanchez1242507}.

Eq. (\ref{eq:RPAcorrelation}) extends the results of previous work
\cite{Ring80,Furche08114105} to fractional systems. Eq. (\ref{eq:RPAEcorrelation})
clarifies the issue on the proper factor for the RPAE correlation
energy. The present derivation shows that the additional factor of
$\frac{1}{2}$ is needed for RPAE correlation energy compared with
that for RPA \cite{Ren127447,Eshuis121084}. Both Eq. (\ref{eq:RPAcorrelation})
and Eq. (\ref{eq:RPAEcorrelation}) apply for both the normal RPA
and also the RPA with fractionally occupied orbitals.

\section{Conclusions}

Many approximate approaches and many-body theoretical methods are
explicit functionals of the single particle Green's function of the
non-interacting reference system. This work shows the rigorous extension
of those methods to fractional charges and fractional spins. This
is achieved by taking the appropriate ensemble average of the one-electron
Green's function, which is the basic variable, and using this in the
corresponding energy expression. We have shown this for methods such
as LDA, HF, MP2 and RPA using the fact that the non-interacting one-electron
density matrix Eq. (\ref{eq:DensityMatrix}) and the non-interacting
polarization propagator Eq. (\ref{eq:pi_0_E_frac}) can be expressed
in terms of the non-interacting single-particle Green's function Eq
(\ref{eq:G_0_E_frac}). This leads to expressions in terms of fractional
charges and fractional spins describing systems that correspond to
the dissociation limit of molecules with integer occupation numbers,
as has been shown for H$_{2}^{+}$ and H$_{2}$ \cite{Cohen08792}.

It should be noted that for many-body methods such as RPA, our development
is different from the finite temperature extensions previously considered
in the literature. The dimension and structure of our RPA matrix equations,
Eq. (\ref{eq:RPA_eig_frac}), are uniquely related to fractional systems
at zero temperature. The development in this paper is completely consistent
with the simple scaling of the orbitals: $\tilde{\phi}_{i}=\sqrt{n_{i}}\phi_{i}$
for the occupied orbitals, and $\tilde{\phi}_{a}=\sqrt{(1-n_{a})}\phi_{a}$
for the unoccupied orbitals, and the inclusion of fractional orbitals
in both the occupied and unoccupied sets of orbitals. For other methods
where the connection to the underlying single particle Green's function
is not clear, we expect the occupation scaling to apply. This development
should allow examining and developing functionals based on many-body
methods to meet the the very challenging exact conditions for fractional
charges and fractional spins \cite{Mori-Sanchez0966403}, which is important,
as the violation of these conditions explains many dramatic failures
of DFT in realistic applications.
\begin{acknowledgments}
Financial support from the Naval Research Office (N00014-09-0576)
(WY) and the National Science Foundation (CHE-09-11119) (WY), the
Royal Society (AJC), Ramón y Cajal and the Spanish Ministry of Science
(FIS2009-12721) (PMS) is gratefully appreciated.
\end{acknowledgments}

\appendix

\section{Details of Derivation}

\subsection{An identity}

Considering
\begin{equation}
\frac{1}{E\pm i\eta}=P\frac{1}{E}\mp i\pi\delta(E)
\end{equation}
then

\begin{eqnarray}
\mathrm{Im}G^{N}(i,j;E)=-\pi\sum_{m}\Big\langle\Psi_{0}^{N}\Big|a_{i}\Big|\Psi_{m}^{N+1}\Big\rangle\Big\langle\Psi_{m}^{N+1}\Big|a_{j}^{\dagger}\Big|\Psi_{0}^{N}\Big\rangle\delta(E-(E_{m}^{N+1}-E_{0}^{N}))\nonumber \\
+\pi\sum_{n}\Big\langle\Psi_{0}^{N}\Big|a_{j}^{\dagger}\Big|\Psi_{n}^{N-1}\Big\rangle\Big\langle\Psi_{n}^{N-1}\Big|a_{i}\Big|\Psi_{0}^{N}\Big\rangle\delta(E-(E_{0}^{N}-E_{n}^{N-1})).
\end{eqnarray}

Let
\begin{equation}
t^{+}=t+\eta,
\end{equation}
 where the infinitesimal number $\eta>0$, then
\begin{eqnarray}
\frac{\partial}{\partial t}G^{N}(i,j;t-t^{+}) & = & i\sum_{n}e^{i(E_{0}^{N}-E_{n}^{N-1})(t^{+}-t)}(-i)(E_{0}^{N}-E_{n}^{N-1})\Big\langle\Psi_{0}^{N}\Big|a_{j}^{\dagger}\Big|\Psi_{n}^{N-1}\Big\rangle\Big\langle\Psi_{n}^{N-1}\Big|a_{i}\Big|\Psi_{0}^{N}\Big\rangle\nonumber \\
 & = & \sum_{n}(E_{0}^{N}-E_{n}^{N-1})\Big\langle\Psi_{0}^{N}\Big|a_{j}^{\dagger}\Big|\Psi_{n}^{N-1}\Big\rangle\Big\langle\Psi_{n}^{N-1}\Big|a_{i}\Big|\Psi_{0}^{N}\Big\rangle
\end{eqnarray}
Thus
\begin{eqnarray}
\int_{C\uparrow}\frac{dE}{2\pi i}EG^{N}(i,j;E) & = & \int_{-\infty}^{+\infty}\frac{dE}{2\pi i}e^{iE\eta}E\left\{ \sum_{m}\frac{\Big\langle\Psi_{0}^{N}\Big|a_{i}\Big|\Psi_{m}^{N+1}\Big\rangle\Big\langle\Psi_{m}^{N+1}\Big|a_{j}^{\dagger}\Big|\Psi_{0}^{N}\Big\rangle}{E-(E_{m}^{N+1}-E_{0}^{N})+i\eta}\right.\nonumber \\
 &  & \left.+\sum_{n}\frac{\Big\langle\Psi_{0}^{N}\Big|a_{j}^{\dagger}\Big|\Psi_{n}^{N-1}\Big\rangle\Big\langle\Psi_{n}^{N-1}\Big|a_{i}\Big|\Psi_{0}^{N}\Big\rangle}{E-(E_{0}^{N}-E_{n}^{N-1})-i\eta}\right\} \nonumber \\
 & = & \sum_{n}(E_{0}^{N}-E_{n}^{N-1})\Big\langle\Psi_{0}^{N}\Big|a_{j}^{\dagger}\Big|\Psi_{n}^{N-1}\Big\rangle\Big\langle\Psi_{n}^{N-1}\Big|a_{i}\Big|\Psi_{0}^{N}\Big\rangle\nonumber \\
 & = & \frac{\partial}{\partial t}G^{N}(i,j;t,t^{+})\nonumber \\
 & = & \frac{1}{\pi}\intop_{-\infty}^{\varepsilon_{F}^{-}}dE\, E\mathrm{Im}G^{N}(i,j;E)\label{eq:Identity}
\end{eqnarray}
 where $\varepsilon_{F}^{-}=E_{0}^{N}-E_{0}^{N-1}=\mu^{-}$ , $\varepsilon_{F}^{+}=E_{0}^{N+1}-E_{0}^{N}=\mu^{+}$.

\subsection{The second order self energy\label{sub:The-second-order}}

The second order self energy is given in terms of $G^{0}$ and we
perform the energy integration

\begin{eqnarray*}
\Sigma^{*(2)}(i,j,E) & = & -\frac{1}{2}\int_{-\infty}^{+\infty}\frac{dE_{1}}{2\pi i}\int_{-\infty}^{+\infty}\frac{dE_{2}}{2\pi i}\\
 &  & \sum_{klm}\sum_{npq}\langle ik||lm\rangle\langle np||jq\rangle G^{0}(l,n;E_{1})G^{0}(m,p;E_{2})G^{0}(q,k;E_{1}+E_{2}-E)\\
 & = & -\frac{1}{2}\sum_{klm}\sum_{npq}\langle ik||lm\rangle\langle np||jq\rangle\\
 &  & \int_{-\infty}^{+\infty}\frac{dE_{1}}{2\pi i}G^{o}(l,n;E_{1})\delta_{mp}\delta_{qk}\left\{ \frac{(1-n_{m})n_{q}}{E-E_{1}+\varepsilon_{q}-\varepsilon_{m}+i\eta}-\frac{n_{q}(1-n_{m})}{E-E_{1}+\varepsilon_{q}-\varepsilon_{m}-i\eta}\right\} \\
 & = & -\frac{1}{2}\sum_{klm}\sum_{npq}\langle ik||lm\rangle\langle np||jq\rangle\delta_{mp}\delta_{qk}\delta_{ln}\\
 &  & \left\{ -\frac{(1-n_{l})(1-n_{m})n_{q}}{E-\varepsilon_{l}-\varepsilon_{m}+\varepsilon_{q}+i\eta}-\frac{n_{l}n_{m}(1-n_{q})}{E-\varepsilon_{l}-\varepsilon_{m}+\varepsilon_{q}-i\eta}\right\} \\
 & = & \frac{1}{2}\sum_{lmq}\langle iq||lm\rangle\langle lm||jq\rangle\left\{ \frac{(1-n_{l})(1-n_{m})n_{q}}{E-\varepsilon_{l}-\varepsilon_{m}+\varepsilon_{q}+i\eta}+\frac{n_{l}n_{m}(1-n_{q})}{E-\varepsilon_{l}-\varepsilon_{m}+\varepsilon_{q}-i\eta}\right\} ,
\end{eqnarray*}
which is Eq. (\ref{eq:SelfEnergy2}) in the text.

We have used the following integrals

\begin{eqnarray}
I & = & \int_{-\infty}^{+\infty}\frac{dE'}{2\pi i}\left\{ \frac{H_{1}}{E'-h_{1}+i\eta}+\frac{P_{1}}{E'-p_{1}-i\eta}\right\} \left\{ \frac{H_{2}}{E'-E-h_{2}+i\eta}+\frac{P_{2}}{E'-E-p_{2}-i\eta}\right\} \nonumber \\
 & = & \mbox{}\frac{H_{1}P_{2}}{E+p_{2}-h_{1}+i\eta}-\frac{H_{2}P_{1}}{E+h_{2}-p_{1}-i\eta},
\end{eqnarray}
such that

\begin{eqnarray}
I_{1} & = & \int_{-\infty}^{+\infty}\frac{dE_{2}}{2\pi i}\left\{ \frac{(1-n_{m})}{E_{2}-\varepsilon_{m}+i\eta}+\frac{n_{m}}{E_{2}-\varepsilon_{m}-i\eta}\right\} \left\{ \frac{(1-n_{q})}{E_{1}+E_{2}-E-\varepsilon_{q}+i\eta}+\frac{n_{q}}{E_{1}+E_{2}-E-\varepsilon_{q}-i\eta}\right\} \nonumber \\
 & = & \frac{(1-n_{m})n_{q}}{E-E_{1}+\varepsilon_{q}-\varepsilon_{m}+i\eta}-\frac{n_{m}(1-n_{q})}{E-E_{1}+\varepsilon_{q}-\varepsilon_{m}-i\eta},
\end{eqnarray}
and

\begin{eqnarray}
I_{2} & = & \int_{-\infty}^{+\infty}\frac{dE_{1}}{2\pi i}\left\{ \frac{(1-n_{l})}{E_{1}-\varepsilon_{l}+i\eta}+\frac{n_{l}}{E_{1}-\varepsilon_{l}-i\eta}\right\} \left\{ \frac{(1-n_{m})n_{q}}{E-E_{1}+\varepsilon_{q}-\varepsilon_{m}+i\eta}-\frac{n_{m}(1-n_{q})}{E-E_{1}+\varepsilon_{q}-\varepsilon_{m}-i\eta}\right\} \nonumber \\
 & = & \int_{-\infty}^{+\infty}\frac{dE_{1}}{2\pi i}\left\{ \frac{(1-n_{l})}{E_{1}-\varepsilon_{l}+i\eta}+\frac{n_{l}}{E_{1}-\varepsilon_{l}-i\eta}\right\} \left\{ -\frac{(1-n_{m})n_{q}}{E_{1}-E-\varepsilon_{q}+\varepsilon_{m}-i\eta}+\frac{n_{m}(1-n_{q})}{E_{1}-E-\varepsilon_{q}+\varepsilon_{m}+i\eta}\right\} \nonumber \\
 & = & \frac{(1-n_{l})(1-n_{m})n_{q}}{\varepsilon_{l}-E-\varepsilon_{q}+\varepsilon_{m}-i\eta}+\frac{n_{l}n_{m}(1-n_{q})}{\varepsilon_{l}-E-\varepsilon_{q}+\varepsilon_{m}+i\eta}\nonumber \\
 & = & -\frac{(1-n_{l})(1-n_{m})n_{q}}{E-\varepsilon_{l}-\varepsilon_{m}+\varepsilon_{q}+i\eta}-\frac{n_{l}n_{m}(1-n_{q})}{E-\varepsilon_{l}-\varepsilon_{m}+\varepsilon_{q}-i\eta}.
\end{eqnarray}

\subsection{Energy Expression from the equation of motion\label{sub:Use-of-theEOM}}

The equation of motion for $H(\lambda)$, suppressing the index of
$\lambda$ in $a_{i}(t)$, can be expressed as
\begin{eqnarray}
i\frac{\partial}{\partial t}a_{i}(t) & = & [a_{i}(t),H(\lambda)]\nonumber \\
 & = & \sum_{j}\left[h_{ij}+(1-\lambda)u_{ij}\right]a_{j}(t)+\lambda\sum_{jkl}\langle ij|kl\rangle a_{j}^{\dagger}(t)a_{l}(t)a_{k}(t).
\end{eqnarray}
Then

\begin{equation}
\lambda\sum_{jkl}\langle ij|kl\rangle a_{j}^{\dagger}(t)a_{l}(t)a_{k}(t)=i\frac{\partial}{\partial t}a_{i}(t)-\sum_{j}\left[h_{ij}+(1-\lambda)u_{ij}\right]a_{j}(t),
\end{equation}
and (using $t^{+}=t+0$)
\begin{equation}
\lambda\sum_{ijkl}\langle ij|kl\rangle a_{i}^{\dagger}(t^{+})a_{j}^{\dagger}(t)a_{l}(t)a_{k}(t)=\sum_{i}a_{i}^{\dagger}(t^{+})\left(i\frac{\partial}{\partial t}\right)a_{i}(t)-\sum_{ij}\left[h_{ij}+(1-\lambda)u_{ij}\right]a_{i}^{\dagger}(t^{+})a_{j}(t)
\end{equation}
 Thus,
\begin{eqnarray*}
\left\langle \Psi_{0}^{N}(\lambda)\left|\lambda V\right|\Psi_{0}^{N}(\lambda)\right\rangle  & = & \frac{1}{2}\lambda\sum_{ijkl}\langle ij|kl\rangle\left\langle \Psi_{0}^{N}(\lambda)\left|a_{i}^{\dagger}(t)a_{j}^{\dagger}(t)a_{l}(t)a_{k}(t)\right|\Psi_{0}^{N}(\lambda)\right\rangle \\
 & = & \frac{1}{2}\sum_{i}\left\langle \Psi_{0}^{N}(\lambda)\left|a_{i}^{\dagger}(t^{+})\left(i\frac{\partial}{\partial t}\right)a_{i}(t)\right|\Psi_{0}^{N}(\lambda)\right\rangle \\
 &  & -\frac{1}{2}\sum_{ij}\left[h_{ij}+(1-\lambda)u_{ij}\right]\left\langle \Psi_{0}^{N}(\lambda)\left|a_{i}^{\dagger}(t^{+})a_{j}(t)\right|\Psi_{0}^{N}(\lambda)\right\rangle \\
 & = & \frac{1}{2}\sum_{i}\left[\left(i\frac{\partial}{\partial t}\right)\left\langle \Psi_{0}^{N}(\lambda)\left|a_{i}^{\dagger}(t^{+})a_{i}(t)\right|\Psi_{0}^{N}(\lambda)\right\rangle \right]\\
 &  & -\frac{1}{2}\sum_{ij}\left[h_{ij}+(1-\lambda)u_{ij}\right]G^{\lambda}(j,i;t-t^{+})\\
 & = & \frac{1}{2}\mathrm{Tr}\left(i\frac{\partial}{\partial t}\right)\mathbf{G}^{\lambda}(t-t^{+})-\mathrm{\frac{1}{2}Tr}\left[\mathbf{h}+\mathbf{u}-\lambda\mathbf{u}\right]\mathbf{G}^{\lambda}(t-t^{+})\\
 & = & \frac{1}{2}\mathrm{Tr}\left[i\frac{\partial}{\partial t}-(\mathbf{h}+\mathbf{u}-\lambda\mathbf{u})\right]\mathbf{G}^{\lambda}(t,t^{+})\\
 & = & \frac{1}{2}\int_{C\uparrow}\frac{dE}{2\pi i}\mathrm{Tr}\left\{ \left[E-(\mathbf{h}+\mathbf{u}-\lambda\mathbf{u})\right]\mathbf{G}^{\lambda}(E)\right\}
\end{eqnarray*}
 using
\begin{eqnarray}
G^{\lambda}(i,j;t-t^{+}) & = & \int_{C\uparrow}\frac{dE}{2\pi i}G^{\lambda}(i,j;E)\nonumber \\
 & = & \int_{C\uparrow}\frac{dE}{2\pi i}\int_{-\infty}^{+\infty}d(t-t^{+})e^{iE(t-t^{+})}G^{\lambda}(i,j;t-t^{+})
\end{eqnarray}
 and the identity of Eq. (\ref{eq:Identity}).

Therefore at $\lambda=1$,
\begin{eqnarray}
\left\langle \Psi_{0}^{N}(1)\left|V\right|\Psi_{0}^{N}(1)\right\rangle  & = & \frac{1}{2}\int_{C\uparrow}\frac{dE}{2\pi i}\mathrm{Tr}\left\{ \left[E-\mathbf{h}\right]\mathbf{G}^{\lambda=1}(E)\right\} \label{eq:V1}
\end{eqnarray}
 and similarly
\begin{equation}
\left\langle \Psi_{0}^{N}(1)\left|h\right|\Psi_{0}^{N}(1)\right\rangle =\int_{C\uparrow}\frac{dE}{2\pi i}\mathrm{Tr}\left\{ \mathbf{h}\mathbf{G}^{\lambda=1}(E)\right\}
\end{equation}

Thus, the ground state energy can be expressed as
\begin{eqnarray*}
E_{0}^{N} & = & E(1)\\
 & = & \left\langle \Psi_{0}^{N}(1)\left|h\right|\Psi_{0}^{N}(1)\right\rangle +\left\langle \Psi_{0}^{N}(1)\left|V\right|\Psi_{0}^{N}(1)\right\rangle \\
 & = & \int_{C\uparrow}\frac{dE}{2\pi i}\mathrm{Tr}\left\{ \mathbf{h}\mathbf{G}^{\lambda=1}(E)\right\} +\frac{1}{2}\int_{C\uparrow}\frac{dE}{2\pi i}\mathrm{Tr}\left\{ \left[E-\mathbf{h}\right]\mathbf{G}^{\lambda=1}(E)\right\} \\
 & = & \frac{1}{2}\int_{C\uparrow}\frac{dE}{2\pi i}\mathrm{Tr}\left\{ \left(E+\mathbf{h}\right)\mathbf{G}(E)\right\}
\end{eqnarray*}

Taking the choice of $\mathbf{h}+\mathbf{u}$ as the Hamiltonian for
the non-interacting reference system,
\begin{equation}
\left[i\frac{\partial}{\partial t}-(\mathbf{h}+\mathbf{u)}\right]\mathbf{G}^{0}(t)=\mathbf{I},
\end{equation}
we have
\begin{equation}
\left[i\frac{\partial}{\partial t}-(\mathbf{h}+\mathbf{u}+\mathbf{\Sigma^{*\lambda}})\right]\mathbf{G}^{\lambda}(t)=\mathbf{I},
\end{equation}
and
\begin{equation}
\left[E-(\mathbf{h}+\mathbf{u}+\mathbf{\Sigma^{*\lambda}})\right]\mathbf{G}^{\lambda}(E)=\mathbf{I},
\end{equation}

then
\begin{eqnarray}
\left(\mathbf{G}^{\lambda}\right)^{-1}(t) & = & \left[i\frac{\partial}{\partial t}-(\mathbf{h}+\mathbf{u}+\Sigma^{*\lambda})\right]\nonumber \\
 & = & \left(\mathbf{G}^{0}\right)^{-1}(t)+\Sigma^{*\lambda}.
\end{eqnarray}
 This gives

\begin{equation}
\left[E-(\mathbf{h}+\mathbf{u}-\lambda\mathbf{u})\right]\mathbf{G}^{\lambda}(E)=\mathbf{I}+\left(\mathbf{\Sigma^{*\lambda}+\lambda u}\right)\mathbf{G}^{\lambda}(E)
\end{equation}

The expression for the energy can be cast as, using Eqs. (\ref{eq:TotalE0}
and \ref{eq:V1} )
\begin{eqnarray*}
E(1)-E(0) & = & \intop_{0}^{1}d\lambda\left\langle \Psi_{0}^{N}(\lambda)\left|H_{1}\right|\Psi_{0}^{N}(\lambda)\right\rangle \\
 & = & \intop_{0}^{1}d\lambda\left[-\sum_{ij}u_{ij}\left\langle \Psi_{0}^{N}(\lambda)\left|a_{i}^{\dagger}a_{j}\right|\Psi_{0}^{N}(\lambda)\right\rangle \right.\\
 &  & \left.+\frac{1}{2}\sum_{ijkl}\langle ij|kl\rangle\left\langle \Psi_{0}^{N}(\lambda)\left|a_{i}^{\dagger}a_{j}^{\dagger}a_{l}a_{k}\right|\Psi_{0}^{N}(\lambda)\right\rangle \right]\\
 & = & \intop_{0}^{1}d\lambda\left[-\sum_{ij}u_{ij}\left\langle \Psi_{0}^{N}(\lambda)\left|a_{i}^{\dagger}a_{j}\right|\Psi_{0}^{N}(\lambda)\right\rangle \right.\\
 &  & \left.+\frac{1}{2}\frac{1}{\lambda}\sum_{ij}\left\langle \Psi_{0}^{N}(\lambda)\left|a_{i}^{+}(t)\left(i\frac{\partial}{\partial t}a_{i}(t)-\sum\left(h_{ij}+u_{ij}-\lambda u_{ij}\right)a_{j}(t)\right)\right|\Psi_{0}^{N}(\lambda)\right\rangle \right]\\
 & = & -\intop_{0}^{1}d\lambda\int_{-\infty}^{+\infty}\frac{dE}{2\pi i}e^{iE\eta}\mathrm{Tr}\left[\mathbf{u}\mathbf{G}^{\lambda}(E)\right]\\
 &  & +\frac{1}{2}\intop_{0}^{1}\frac{d\lambda}{\lambda}\int_{-\infty}^{+\infty}\frac{dE}{2\pi i}e^{iE\eta}\mathrm{Tr}\left[\mathbf{I}+\left(\mathbf{\Sigma}^{*\lambda}(E)+\lambda\mathbf{u}\right)\mathbf{G}^{\lambda}(E)\right]\\
 & = & -\intop_{0}^{1}d\lambda\int_{-\infty}^{+\infty}\frac{dE}{2\pi i}e^{iE\eta}\mathrm{Tr}\mathbf{u}\mathbf{G}^{\lambda}(E)\\
 &  & +\frac{1}{2}\intop_{0}^{1}\frac{d\lambda}{\lambda}\int_{-\infty}^{+\infty}\frac{dE}{2\pi i}e^{iE\eta}\mathrm{Tr}\left(\mathbf{\Sigma}^{*\lambda}(E)+\lambda\mathbf{u}\right)\mathbf{G}^{\lambda}(E)\\
 & = & \frac{1}{2}\intop_{0}^{1}d\lambda\int_{-\infty}^{+\infty}\frac{dE}{2\pi i}e^{iE\eta}\mathrm{Tr}\left[\left(-\mathbf{u}+\frac{1}{\lambda}\mathbf{\Sigma}^{*\lambda}(E)\right)\mathbf{G}^{\lambda}(E)\right],
\end{eqnarray*}
which is Eq. (\ref{eq:TotelE}) in the text.

\subsection{The second-order energy \label{sub:The-second-order-energy}}

The detailed integration needed for the second order energy is the
following

\begin{eqnarray}
 &  & \frac{1}{4}\int_{-\infty}^{+\infty}\frac{dE}{2\pi i}e^{iE\eta}\mathrm{Tr}\{\mathbf{\Sigma}^{*(2)}(E)\mathbf{G}^{0}(E)\}\nonumber \\
 & = & \frac{1}{4}\int_{-\infty}^{+\infty}\frac{dE}{2\pi i}e^{iE\eta}\frac{1}{2}\sum_{ilmq}\langle iq||lm\rangle\langle lm||iq\rangle\nonumber \\
 &  & \left\{ \frac{(1-n_{l})(1-n_{m})n_{q}}{E-\varepsilon_{l}-\varepsilon_{m}+\varepsilon_{q}+i\eta}+\frac{n_{l}n_{m}(1-n_{q})}{E-\varepsilon_{l}-\varepsilon_{m}+\varepsilon_{q}-i\eta}\right\} \left\{ \frac{(1-n_{i})}{E-\varepsilon_{i}+i\eta}+\frac{n_{i}}{E-\varepsilon_{i}-i\eta}\right\} \nonumber \\
 & = & \frac{1}{8}\sum_{ilmq}\langle iq||lm\rangle\langle lm||iq\rangle\left[n_{l}n_{m}(1-n_{q})\left\{ \frac{(1-n_{i})}{-(-\varepsilon_{l}-\varepsilon_{m}+\varepsilon_{q})-\varepsilon_{i}+i\eta}+\frac{n_{i}}{-(-\varepsilon_{l}-\varepsilon_{m}+\varepsilon_{q})-\varepsilon_{i}-i\eta}\right\} \right.\nonumber \\
 &  & \left.+n_{i}\left\{ \frac{(1-n_{l})(1-n_{m})n_{q}}{\varepsilon_{i}-\varepsilon_{l}-\varepsilon_{m}+\varepsilon_{q}+i\eta}+\frac{n_{l}n_{m}(1-n_{q})}{\varepsilon_{i}-\varepsilon_{l}-\varepsilon_{m}+\varepsilon_{q}-i\eta}\right\} \right]\nonumber \\
 & = & \frac{1}{8}\sum_{ilmq}\langle iq||lm\rangle\langle lm||iq\rangle\left[n_{l}n_{m}(1-n_{q})\left\{ \frac{1}{\varepsilon_{l}+\varepsilon_{m}-\varepsilon_{q}-\varepsilon_{i}+i\eta}+\frac{n_{i}}{\varepsilon_{i}-\varepsilon_{l}-\varepsilon_{m}+\varepsilon_{q}-i\eta}\right\} \right.\nonumber \\
 &  & \left.+\frac{(1-n_{l})(1-n_{m})n_{q}n_{i}}{\varepsilon_{i}-\varepsilon_{l}-\varepsilon_{m}+\varepsilon_{q}+i\eta}\right]\nonumber \\
 & = & \frac{1}{8}\sum_{ilmq}\langle iq||lm\rangle\langle lm||iq\rangle\left\{ \frac{n_{l}n_{m}(1-n_{q})(1-n_{i)}}{\varepsilon_{l}+\varepsilon_{m}-\varepsilon_{q}-\varepsilon_{i}}+\frac{(1-n_{l})(1-n_{m})n_{q}n_{i}}{\varepsilon_{i}+\varepsilon_{q}-\varepsilon_{l}-\varepsilon_{m}}\right\} \nonumber \\
 & = & \frac{1}{4}\sum_{ilmq}\langle iq||lm\rangle\langle lm||iq\rangle\frac{(1-n_{l})(1-n_{m})n_{q}n_{i}}{\varepsilon_{i}+\varepsilon_{q}-\varepsilon_{l}-\varepsilon_{m}}\label{eq:2ndOrderE}
\end{eqnarray}

\subsection{The integration of the polarization propagator \label{sub:The-integration-of_PI}}

\begin{eqnarray*}
\frac{-1}{\pi}\int_{-\infty}^{\infty}dE\textrm{Im}\Pi(i,j;k,l;E) & = & \sum_{n\neq0}\left[\left\langle \Psi_{0}^{N}\left|a_{j}^{\dagger}a_{i}\left|\Psi_{n}^{N}\right\rangle \left\langle \Psi_{n}^{N}\right|a_{k}^{\dagger}a_{l}\right|\Psi_{0}^{N}\right\rangle +\left\langle \Psi_{0}^{N}\left|a_{k}^{\dagger}a_{l}\left|\Psi_{n}^{N}\right\rangle \left\langle \Psi_{n}^{N}\right|a_{j}^{\dagger}a_{i}\right|\Psi_{0}^{N}\right\rangle \right]\\
 & = & -\left\langle \Psi_{0}^{N}\left|a_{j}^{\dagger}a_{i}\left|\Psi_{0}^{N}\right\rangle \left\langle \Psi_{0}^{N}\right|a_{k}^{\dagger}a_{l}\right|\Psi_{0}^{N}\right\rangle +\sum_{n}\left\langle \Psi_{0}^{N}\left|a_{j}^{\dagger}a_{i}\left|\Psi_{n}^{N}\right\rangle \left\langle \Psi_{n}^{N}\right|a_{k}^{\dagger}a_{l}\right|\Psi_{0}^{N}\right\rangle \\
 &  & -\left\langle \Psi_{0}^{N}\left|a_{k}^{\dagger}a_{l}\left|\Psi_{0}^{N}\right\rangle \left\langle \Psi_{0}^{N}\right|a_{j}^{\dagger}a_{i}\right|\Psi_{0}^{N}\right\rangle +\sum_{n}\left\langle \Psi_{0}^{N}\left|a_{k}^{\dagger}a_{l}\left|\Psi_{n}^{N}\right\rangle \left\langle \Psi_{n}^{N}\right|a_{j}^{\dagger}a_{i}\right|\Psi_{0}^{N}\right\rangle \\
 & = & -\rho_{ji}\rho_{kl}+\left\langle \Psi_{0}^{N}\left|a_{j}^{\dagger}a_{i}a_{k}^{\dagger}a_{l}\right|\Psi_{0}^{N}\right\rangle \\
 &  & -\rho_{kl}\rho_{ji}+\left\langle \Psi_{0}^{N}\left|a_{k}^{\dagger}a_{l}a_{j}^{\dagger}a_{i}\right|\Psi_{0}^{N}\right\rangle \\
 & = & \delta_{ki}\left\langle \Psi_{0}^{N}\left|a_{j}^{\dagger}a_{l}\right|\Psi_{0}^{N}\right\rangle -\left\langle \Psi_{0}^{N}\left|a_{j}^{\dagger}a_{k}^{\dagger}a_{i}a_{l}\right|\Psi_{0}^{N}\right\rangle -\rho_{ji}\rho_{kl}\\
 &  & +\delta_{lj}\left\langle \Psi_{0}^{N}\left|a_{k}^{\dagger}a_{i}\right|\Psi_{0}^{N}\right\rangle -\left\langle \Psi_{0}^{N}\left|a_{k}^{\dagger}a_{j}^{\dagger}a_{l}a_{i}\right|\Psi_{0}^{N}\right\rangle -\rho_{kl}\rho_{ji}\\
 & = & \delta_{ki}\rho_{jl}-\left\langle \Psi_{0}^{N}\left|a_{j}^{\dagger}a_{k}^{\dagger}a_{i}a_{l}\right|\Psi_{0}^{N}\right\rangle -\rho_{ji}\rho_{kl}+\delta_{lj}\rho_{ki}-\left\langle \Psi_{0}^{N}\left|a_{k}^{\dagger}a_{j}^{\dagger}a_{l}a_{i}\right|\Psi_{0}^{N}\right\rangle -\rho_{kl}\rho_{ji}\\
 & = & \delta_{ki}\rho_{jl}-2\gamma_{jk,li}-\rho_{ji}\rho_{kl}+\delta_{lj}\rho_{ki}-2\gamma_{kj,il}-\rho_{kl}\rho_{ji}\\
 & = & \delta_{ki}\rho_{jl}+\delta_{lj}\rho_{ki}-2\rho_{kl}\rho_{ji}-4\gamma_{jk,li}\\
 & = & \delta_{ki}\rho_{jl}+\left\langle \Psi_{0}^{N}\left|a_{j}^{\dagger}a_{k}^{\dagger}a_{l}a_{i}\right|\Psi_{0}^{N}\right\rangle -\rho_{ji}\rho_{kl}+\delta_{lj}\rho_{ki}+\left\langle \Psi_{0}^{N}\left|a_{k}^{\dagger}a_{j}^{\dagger}a_{i}a_{l}\right|\Psi_{0}^{N}\right\rangle -\rho_{kl}\rho_{ji}\\
 & = & \delta_{ki}\rho_{jl}+2\gamma_{jk,il}-\rho_{ji}\rho_{kl}+\delta_{lj}\rho_{ki}+2\gamma_{kj,li}-\rho_{kl}\rho_{ji}\\
 & = & \delta_{ki}\rho_{jl}+\delta_{lj}\rho_{ki}-2\rho_{kl}\rho_{ji}+4\gamma_{jk,il},
\end{eqnarray*}
where we used
\[
2\gamma_{jk,li}=2\gamma_{kj,il}=\left\langle \Psi_{0}^{N}\left|a_{j}^{\dagger}a_{k}^{\dagger}a_{i}a_{l}\right|\Psi_{0}^{N}\right\rangle
\]
and

\[
2\gamma_{jk,il}=-2\gamma_{jk,li}=\left\langle \Psi_{0}^{N}\left|a_{j}^{\dagger}a_{k}^{\dagger}a_{l}a_{i}\right|\Psi_{0}^{N}\right\rangle .
\]

\subsection{The details of the potential energy expression \label{sub:The-details-of_P_E}}

\begin{eqnarray*}
 &  & \frac{1}{2}\sum_{ijkl}\langle ij|\bar{V}_{ph}|kl\rangle\left[\rho_{kl}\rho_{ji}-\frac{1}{2}\left(\delta_{ki}\rho_{jl}+\delta_{lj}\rho_{ki}\right)\right]\\
 & = & \frac{1}{2}\sum_{ijkl}\int\int d\mathbf{x}_{1}d\mathbf{x}_{2}\phi_{i}^{*}(\mathbf{x}_{1})\phi_{l}^{*}(\mathbf{x}_{2})v(\mathbf{x}_{1},\mathbf{x}_{2})\phi_{j}(\mathbf{x}_{1})\phi_{k}(\mathbf{x}_{2})\left[\rho_{kl}\rho_{ji}-\frac{1}{2}\left(\delta_{ki}\rho_{jl}+\delta_{lj}\rho_{ki}\right)\right]\\
 & = & \frac{1}{2}\int\int d\mathbf{x}_{1}d\mathbf{x}_{2}v(\mathbf{x}_{1},\mathbf{x}_{2})\rho(\mathbf{x}_{1},\mathbf{x}_{1})\rho(\mathbf{x}_{2},\mathbf{x}_{2})\\
 &  & -\frac{1}{4}\int\int d\mathbf{x}_{1}d\mathbf{x}_{2}v(\mathbf{x}_{1},\mathbf{x}_{2})\left(\rho(\mathbf{x}_{1},\mathbf{x}_{2})\delta(\mathbf{x}_{2}-\mathbf{x}_{1})+\rho(\mathbf{x}_{2},\mathbf{x}_{1})\delta(\mathbf{x}_{2}-\mathbf{x}_{1})\right)\\
 & = & \frac{1}{2}\int\int d\mathbf{x}_{1}d\mathbf{x}_{2}v(\mathbf{x}_{1},\mathbf{x}_{2})\left[\rho(\mathbf{x}_{1})\rho(\mathbf{x}_{2})-\rho(\mathbf{x}_{1})\delta(\mathbf{x}_{2}-\mathbf{x}_{1})\right]
\end{eqnarray*}
This is just part of Eq. () in the text. And
\begin{eqnarray*}
 &  & \frac{1}{4}\sum_{ijkl}\langle ij|V_{ph}|kl\rangle\left[\rho_{kl}\rho_{ji}-\frac{1}{2}\left(\delta_{ki}\rho_{jl}+\delta_{lj}\rho_{ki}\right)\right]\\
 & = & \frac{1}{4}\sum_{ijkl}\int\int d\mathbf{x}_{1}d\mathbf{x}_{2}\phi_{i}^{*}(\mathbf{x}_{1})\phi_{l}^{*}(\mathbf{x}_{2})v(\mathbf{x}_{1},\mathbf{x}_{2})(1-P_{12})\phi_{j}(\mathbf{x}_{1})\phi_{k}(\mathbf{x}_{2})\left[\rho_{kl}\rho_{ji}-\frac{1}{2}\left(\delta_{ki}\rho_{jl}+\delta_{lj}\rho_{ki}\right)\right]\\
 & = & \frac{1}{4}\int\int d\mathbf{x}_{1}d\mathbf{x}_{2}v(\mathbf{x}_{1},\mathbf{x}_{2})\left(\rho(\mathbf{x}_{1},\mathbf{x}_{1})\rho(\mathbf{x}_{2},\mathbf{x}_{2})-\rho(\mathbf{x}_{2},\mathbf{x}_{1})\rho(\mathbf{x}_{1},\mathbf{x}_{2})\right)\\
 &  & -\frac{1}{8}\int\int d\mathbf{x}_{1}d\mathbf{x}_{2}v(\mathbf{x}_{1},\mathbf{x}_{2})(\rho(\mathbf{x}_{1},\mathbf{x}_{2})\delta(\mathbf{x}_{2}-\mathbf{x}_{1})\\
 &  & -\rho(\mathbf{x}_{2},\mathbf{x}_{2})\delta(\mathbf{x}_{1}-\mathbf{x}_{1})+\rho(\mathbf{x}_{2},\mathbf{x}_{1})\delta(\mathbf{x}_{2}-\mathbf{x}_{1})-\rho(\mathbf{x}_{1},\mathbf{x}_{1})\delta(\mathbf{x}_{2}-\mathbf{x}_{2}))\\
 & = & \frac{1}{4}\int\int d\mathbf{x}_{1}d\mathbf{x}_{2}v(\mathbf{x}_{1},\mathbf{x}_{2})\left(\rho(\mathbf{x}_{1})\rho(\mathbf{x}_{2})-\rho(\mathbf{x}_{2},\mathbf{x}_{1})\rho(\mathbf{x}_{1},\mathbf{x}_{2})\right)\\
 &  & -\frac{1}{8}\int\int d\mathbf{x}_{1}d\mathbf{x}_{2}v(\mathbf{x}_{1},\mathbf{x}_{2})\left(2\rho(\mathbf{x}_{1})\delta(\mathbf{x}_{2}-\mathbf{x}_{1})-\rho(\mathbf{x}_{1})\delta(0)-\rho(\mathbf{x}_{2})\delta(0)\right)\\
 & = & \frac{1}{4}\int\int d\mathbf{x}_{1}d\mathbf{x}_{2}v(\mathbf{x}_{1},\mathbf{x}_{2})\left(\rho(\mathbf{x}_{1})\rho(\mathbf{x}_{2})-\rho(\mathbf{x}_{2},\mathbf{x}_{1})\rho(\mathbf{x}_{1},\mathbf{x}_{2})\right)\\
 &  & -\frac{1}{4}\int\int d\mathbf{x}_{1}d\mathbf{x}_{2}v(\mathbf{x}_{1},\mathbf{x}_{2})\left(\rho(\mathbf{x}_{1})\delta(\mathbf{x}_{2}-\mathbf{x}_{1})-\rho(\mathbf{x}_{1})\delta(0)\right).
\end{eqnarray*}

\subsection{RPA correlation energy\label{sub:RPA-correlation-energy}}

Here we will evaluate the integral of Eq. (\ref{eq:E_integ}), using
some techniques from Ref. (\cite{Blaizot86}). Based on $\lim_{E\rightarrow\infty}\mathbf{\Pi}^{0}(E)\sim1/E$,
\[
\lim_{E\rightarrow\infty}E\frac{d}{dE}\textrm{Tr}\left[\ln(\mathbf{1}-\mathbf{\Pi}^{0}(E)\mathbf{U}_{ph})\right]=\lim_{E\rightarrow-\infty}E\frac{d}{dE}\textrm{Tr}\left[\ln(\mathbf{1}-\mathbf{\Pi}^{0}(E)\mathbf{U}_{ph})\right],
\]
 because $\left[\ln(\mathbf{1}-\mathbf{\Pi}^{0}(E)\mathbf{U}_{ph})\right]=-\mathbf{\Pi}^{0}(E)\mathbf{U}_{ph}-\frac{1}{2}\left(\mathbf{\Pi}^{0}(E)\mathbf{U}_{ph}\right)^{2}-\frac{1}{3}\left(\mathbf{\Pi}^{0}(E)\mathbf{U}_{ph}\right)^{3}+....$.
and $\lim_{E\rightarrow\infty}E\frac{d}{dE}\textrm{Tr}\left[\ln(\mathbf{1}-\mathbf{\Pi}^{0}(E)\mathbf{U}_{ph})\right]$
is an even function of $E$. We thus have

\begin{eqnarray}
 &  & \frac{1}{4\pi}\int_{-\infty}^{\infty}dE\textrm{Im}\textrm{Tr}\left[\ln(\mathbf{1}-\mathbf{\Pi}^{0}(E)\mathbf{U}_{ph})+\mathbf{\Pi}^{0}(E)\mathbf{U}_{ph}\right]\nonumber \\
 & = & -\frac{1}{4\pi}\int_{-\infty}^{\infty}EdE\frac{d}{dE}\textrm{ImTr}\left[\ln(\mathbf{1}-\mathbf{\Pi}^{0}(E)\mathbf{U}_{ph})\right]+\frac{1}{4\pi}\int_{-\infty}^{\infty}dE\textrm{Im}\textrm{Tr}\left[\mathbf{\Pi}^{0}(E)\mathbf{U}_{ph}\right]\nonumber \\
 &  & +\frac{1}{4\pi}\left.E\textrm{ImTr}\left[\ln(\mathbf{1}-\mathbf{\Pi}^{0}(E)\mathbf{U}_{ph})\right]\right|_{-\infty}^{\infty}\nonumber \\
 & = & -\frac{1}{4\pi}\int_{-\infty}^{\infty}EdE\frac{d}{dE}\textrm{ImTr}\left[\ln(\mathbf{1}-\mathbf{\Pi}^{0}(E)\mathbf{U}_{ph})\right]+\frac{1}{4\pi}\int_{-\infty}^{\infty}dE\textrm{Im}\textrm{Tr}\left[\mathbf{\Pi}^{0}(E)\mathbf{U}_{ph}\right],\label{eq:3parts}
\end{eqnarray}
We will consider the two part in Eq (\ref{eq:3parts}) separately.
Using Eq. (\ref{eq:RPAmatrix}) we get
\[
\left(\mathbf{\Pi}^{\mathrm{RPA}}\right)^{-1}=\left(\mathbf{\Pi}^{0}\right)^{-1}-\mathbf{U}_{ph}.
\]
Then

\begin{eqnarray*}
\frac{d}{dE}\textrm{Tr}\left[\ln(\mathbf{1}-\mathbf{\Pi}^{0}(E)\mathbf{U}_{ph})\right] & = & \frac{d}{dE}\textrm{\textrm{Tr}\ensuremath{\left[\ln\left(\mathbf{\Pi}^{0}\left(\mathbf{\Pi}^{RPA}\right)^{-1}\right)\right]}}\\
 & = & \textrm{Tr}\left[\left(\mathbf{\Pi}^{0}\right)^{-1}\frac{d}{dE}\mathbf{\Pi}^{0}+\left(\mathbf{\Pi}^{\mathrm{RPA}}\right)\frac{d}{dE}\left(\mathbf{\Pi}^{\mathrm{RPA}}\right)^{-1}\right]\\
 & = & \textrm{Tr}\left[\left(\mathbf{\Pi}^{0}\right)^{-1}\frac{d}{dE}\mathbf{\Pi}^{0}+\left(\mathbf{\Pi}^{\mathrm{RPA}}\right)\frac{d}{dE}\left(\mathbf{\Pi}^{0}\right)^{-1}\right]
\end{eqnarray*}

\[
\Pi^{0,N+\delta}(i,j;k,l;E)=\delta_{ik}\delta_{jl}\left\{ \frac{(1-n_{i})n_{j}}{E-(\varepsilon_{i}-\varepsilon_{j})+i\eta}-\frac{n_{i}(1-n_{j})}{E+(\varepsilon_{j}-\varepsilon_{i})-i\eta}\right\}
\]

\begin{eqnarray*}
 &  & \textrm{Tr}\left[\left(\mathbf{\Pi}^{0}\right)^{-1}\frac{d}{dE}\mathbf{\Pi}^{0}\right]\\
 & = & \sum_{ij}\left\{ \frac{(1-n_{i})n_{j}}{E-(\varepsilon_{i}-\varepsilon_{j})+i\eta}-\frac{n_{i}(1-n_{j})}{E+(\varepsilon_{j}-\varepsilon_{i})-i\eta}\right\} ^{-1}\frac{d}{dE}\left\{ \frac{(1-n_{i})n_{j}}{E-(\varepsilon_{i}-\varepsilon_{j})+i\eta}-\frac{n_{i}(1-n_{j})}{E+(\varepsilon_{j}-\varepsilon_{i})-i\eta}\right\} \\
 & = & \sum_{ij}\left\{ \frac{(1-n_{i})n_{j}}{E-(\varepsilon_{i}-\varepsilon_{j})+i\eta}-\frac{n_{i}(1-n_{j})}{E+(\varepsilon_{j}-\varepsilon_{i})-i\eta}\right\} ^{-1}\left\{ -\frac{(1-n_{i})n_{j}}{\left(E-(\varepsilon_{i}-\varepsilon_{j})+i\eta\right)^{2}}+\frac{n_{i}(1-n_{j})}{\left(E+(\varepsilon_{j}-\varepsilon_{i})-i\eta\right)^{2}}\right\}
\end{eqnarray*}
\begin{eqnarray*}
 &  & -\frac{1}{4\pi}\int_{-\infty}^{\infty}EdE\textrm{ImTr}\left[\left(\mathbf{\Pi}^{0}\right)^{-1}\frac{d}{dE}\mathbf{\Pi}^{0}\right]\\
 & = & -\frac{1}{4\pi}\int_{-\infty}^{\infty}EdE\sum_{ij}\left\{ \frac{(1-n_{i})n_{j}}{E-(\varepsilon_{i}-\varepsilon_{j})+i\eta}-\frac{n_{i}(1-n_{j})}{E+(\varepsilon_{j}-\varepsilon_{i})-i\eta}\right\} ^{-1}\\
 &  & \left\{ \frac{(1-n_{i})n_{j}}{\left(E-(\varepsilon_{i}-\varepsilon_{j})+i\eta\right)}\pi\delta(E-(\varepsilon_{i}-\varepsilon_{j}))+\frac{n_{i}(1-n_{j})}{\left(E+(\varepsilon_{j}-\varepsilon_{i})-i\eta\right)}\pi\delta(E+(\varepsilon_{j}-\varepsilon_{i}))\right\} \\
 & = & -\frac{1}{4}\left\{ \sum_{ij}(\varepsilon_{i}-\varepsilon_{j})\frac{(1-n_{i})n_{j}}{(1-n_{i})n_{j}-n_{i}(1-n_{j})}+\sum_{ij}(\varepsilon_{i}-\varepsilon_{j})\frac{n_{i}(1-n_{j})}{(1-n_{i})n_{j}-n_{i}(1-n_{j})}\right\} \\
 & = & -\frac{1}{4}\left[\sum_{i\in\mathrm{particle},j\in\mathrm{hole}}(\varepsilon_{i}-\varepsilon_{j})\frac{(1-n_{i})n_{j}}{(1-n_{i})n_{j}-n_{i}(1-n_{j})}+\sum_{i\in\mathrm{hole},j\in\mathrm{particle}}(\varepsilon_{i}-\varepsilon_{j})\frac{n_{i}(1-n_{j})}{(1-n_{i})n_{j}-n_{i}(1-n_{j})}\right]\\
 & = & -\frac{1}{4}\left[\sum_{i\in\mathrm{particle},j\in\mathrm{hole}}\left(\varepsilon_{i}-\varepsilon_{j}\right)-\sum_{i\in\mathrm{hole},j\in\mathrm{particle}}\left(\varepsilon_{i}-\varepsilon_{j}\right)\right]\\
 & = & -\frac{1}{2}\sum_{i\in\mathrm{particle},j\in\mathrm{hole}}\left(\varepsilon_{i}-\varepsilon_{j}\right)
\end{eqnarray*}
Note that $\frac{(1-n_{i})n_{j}}{(1-n_{i})n_{j}-n_{i}(1-n_{j})}=1$
for $i\in\mathrm{particle},j\in\mathrm{hole}$, because $n_{i}(1-n_{j})=0$
, excluding the frac-frac block, which is zero from $\left(\varepsilon_{i}-\varepsilon_{j}\right)$.
And $\frac{n_{i}(1-n_{j})}{(1-n_{i})n_{j}-n_{i}(1-n_{j})}=-1$, for
$i\in\mathrm{hole},j\in\mathrm{particle}$, because $(1-n_{i})n_{j}=0$,
excluding the frac-frac block, which is zero from $\left(\varepsilon_{i}-\varepsilon_{j}\right)$.

Consider now

\begin{eqnarray*}
-\frac{1}{4\pi}\int_{-\infty}^{\infty}EdE\textrm{ImTr}\left(\mathbf{\Pi}^{\mathrm{RPA}}\right)\frac{d}{dE}\left(\mathbf{\Pi}^{0}\right)^{-1}
\end{eqnarray*}

\[
\left(\mathbf{\Pi^{\mathbf{0}}}(i,j;k,l;E)\right)^{-1}=\delta_{ik}\delta_{jl}\left\{ \frac{(1-n_{i})n_{j}}{E-(\varepsilon_{i}-\varepsilon_{j})+i\eta}-\frac{n_{i}(1-n_{j})}{E+(\varepsilon_{j}-\varepsilon_{i})-i\eta}\right\} ^{-1}
\]

For $i\in\mathrm{particle},j\in\mathrm{hole}$,

\begin{eqnarray*}
\frac{d}{dE}\left(\mathbf{\Pi^{\mathbf{0}}}(i,j;k,l;E)\right)^{-1}=\delta_{ik}\delta_{jl}\frac{d}{dE}\left\{ \frac{(1-n_{i})n_{j}}{E-(\varepsilon_{i}-\varepsilon_{j})+i\eta}\right\} ^{-1} & = & \delta_{ik}\delta_{jl}\frac{d}{dE}\frac{E-(\varepsilon_{i}-\varepsilon_{j})}{(1-n_{i})n_{j}}\\
 & = & \delta_{ik}\delta_{jl}\frac{1}{(1-n_{i})n_{j}}
\end{eqnarray*}

For $i\in\mathrm{hole},j\in\mathrm{particle}$,

\begin{eqnarray*}
\frac{d}{dE}\left(\mathbf{\Pi^{\mathbf{0}}}(i,j;k,l;E)\right)^{-1}=\delta_{ik}\delta_{jl}\frac{d}{dE}\left\{ -\frac{n_{i}(1-n_{j})}{E+(\varepsilon_{j}-\varepsilon_{i})-i\eta}\right\} ^{-1} & = & \delta_{ik}\delta_{jl}\frac{d}{dE}(-1)\frac{E+(\varepsilon_{j}-\varepsilon_{i})}{n_{i}(1-n_{j})}\\
 & = & \delta_{ik}\delta_{jl}\frac{-1}{n_{i}(1-n_{j})}
\end{eqnarray*}

Define the diagonal matrix

\[
\mathbf{\nu}=\left(\begin{array}{cc}
\frac{1}{(1-n_{i})n_{j}}\\
 & \frac{-1}{n_{i}(1-n_{j})}
\end{array}\right)
\]

\[
\mathbf{\Pi}^{\mathrm{RPA}}=\sum_{n\neq0}\frac{\left(\begin{array}{c}
\mathbf{X}^{n}\\
\mathbf{Y}^{n}
\end{array}\right)(\mathbf{X}^{n\dagger}\mathbf{Y}^{n\dagger})}{E-\varepsilon_{n}^{\pi}+i\eta}-\sum_{n\neq0}\frac{\left(\left(\begin{array}{c}
\mathbf{Y}^{n}\\
\mathbf{X}^{n}
\end{array}\right)(\mathbf{Y}^{n\dagger}\mathbf{X}^{n\dagger})\right)^{*}}{E+\varepsilon_{n}^{\pi}-i\eta}
\]

\begin{eqnarray*}
\textrm{Im}\mathbf{\Pi}^{\mathrm{RPA}} & = & \sum_{n\neq0}\left(\begin{array}{c}
\mathbf{X}^{n}\\
\mathbf{Y}^{n}
\end{array}\right)(\mathbf{X}^{n\dagger}\mathbf{Y}^{n\dagger})(-\pi)\delta(E-\varepsilon_{n}^{\pi})-\sum_{n\neq0}\left(\left(\begin{array}{c}
\mathbf{Y}^{n}\\
\mathbf{X}^{n}
\end{array}\right)(\mathbf{Y}^{n\dagger}\mathbf{X}^{n\dagger})\right)^{*}\pi\delta(E+\varepsilon_{n}^{\pi})
\end{eqnarray*}

Thus

\begin{eqnarray*}
 &  & -\frac{1}{4\pi}\int_{-\infty}^{\infty}EdE\frac{d}{dE}\textrm{ImTr}\left[\left(\mathbf{\Pi}^{RPA}\right)\frac{d}{dE}\left(\mathbf{\Pi}^{0}\right)^{-1}\right]\\
 & = & -\frac{1}{4\pi}\int_{-\infty}^{\infty}EdE\textrm{Tr}\left[\sum_{n\neq0}\left(\begin{array}{c}
\mathbf{X}^{n}\\
\mathbf{Y}^{n}
\end{array}\right)(\mathbf{X}^{n\dagger}\mathbf{Y}^{n\dagger})(-\pi)\delta(E-\varepsilon_{n}^{\pi})-\sum_{n\neq0}\left(\left(\begin{array}{c}
\mathbf{Y}^{n}\\
\mathbf{X}^{n}
\end{array}\right)(\mathbf{Y}^{n\dagger}\mathbf{X}^{n\dagger})\right)^{*}\pi\delta(E+\varepsilon_{n}^{\pi})\right]\mathbf{\nu}\\
 & = & \frac{1}{4}\sum_{n>0}\left[\varepsilon_{n}^{\pi}\textrm{Tr}(\mathbf{X}^{n\dagger}\mathbf{Y}^{n\dagger})\mathbf{\nu}\left(\begin{array}{c}
\mathbf{X}^{n}\\
\mathbf{Y}^{n}
\end{array}\right)-\varepsilon_{n}^{\pi}\textrm{Tr}(\mathbf{Y}^{n\dagger}\mathbf{X}^{n\dagger})\mathbf{\nu}\left(\begin{array}{c}
\mathbf{Y}^{n}\\
\mathbf{X}^{n}
\end{array}\right)\right]\\
 & = & \frac{1}{2}\sum_{n>0}\varepsilon_{n}^{\pi}
\end{eqnarray*}
where we use the normalization condition for the RPA matrix eigenvalue
problem, Eq. (\ref{eq:RPA_eig_frac})

\[
(\mathbf{X}^{n\dagger}\mathbf{Y}^{n\dagger})\mathbf{\nu}\left(\begin{array}{c}
\mathbf{X}^{n}\\
\mathbf{Y}^{n}
\end{array}\right)=(\mathbf{\tilde{X}}^{n\dagger}\mathbf{\tilde{Y}}^{n\dagger})\left(\begin{array}{c}
\mathbf{\tilde{X}}^{n}\\
\mathbf{-\tilde{Y}}^{n}
\end{array}\right)=1
\]
\[
(\mathbf{Y}^{n\dagger}\mathbf{X}^{n\dagger})\mathbf{\nu}\left(\begin{array}{c}
\mathbf{Y}^{n}\\
\mathbf{X}^{n}
\end{array}\right)=(\mathbf{\tilde{Y}}^{n\dagger}\mathbf{\tilde{X}}^{n\dagger})\left(\begin{array}{c}
\mathbf{\tilde{Y}}^{n}\\
-\mathbf{\tilde{X}}^{n}
\end{array}\right)=-1.
\]

Now consider the last term in Eq. (\ref{eq:3parts}),

\begin{eqnarray*}
 &  & \frac{1}{4\pi}\int_{-\infty}^{\infty}dE\textrm{Im}\textrm{Tr}\left[\mathbf{\Pi}^{0}(E)\mathbf{U}_{ph}\right]\\
 & = & -\frac{1}{4}\textrm{Tr}\left[\left(\begin{array}{c}
\mathbf{\tilde{X}}^{0,n}\\
\mathbf{\tilde{Y}}^{0,n}
\end{array}\right)(\mathbf{\tilde{X}}^{0,n\dagger}\mathbf{\tilde{Y}}^{0,n\dagger})\mathbf{U}_{ph}+\left(\left(\begin{array}{c}
\mathbf{\tilde{Y}}^{0,n}\\
\mathbf{\tilde{X}}^{0,n}
\end{array}\right)(\mathbf{\tilde{Y}}^{0,n\dagger}\mathbf{\tilde{X}}^{0,n\dagger})\right)^{*}\mathbf{U}_{ph}\right]\\
 & = & -\frac{1}{4}\left[\textrm{(\ensuremath{\mathbf{\tilde{X}}^{0,n\dagger}\mathbf{\tilde{Y}}^{0,n\dagger}})}\mathbf{U}_{ph}\left(\begin{array}{c}
\mathbf{\tilde{X}}^{0,n}\\
\mathbf{\tilde{Y}}^{0,n}
\end{array}\right)+\left((\mathbf{\tilde{Y}}^{0,n\dagger}\mathbf{\tilde{X}}^{0,n\dagger})\mathbf{U}_{ph}\left(\begin{array}{c}
\mathbf{\tilde{Y}}^{0,n}\\
\mathbf{\tilde{X}}^{0,n}
\end{array}\right)\right)^{*}\right]\\
 & = & -\frac{1}{2}\textrm{(\ensuremath{\mathbf{\tilde{X}}^{0,n\dagger}\mathbf{\tilde{Y}}^{0,n\dagger}})}\mathbf{U}_{ph}\left(\begin{array}{c}
\mathbf{\tilde{X}}^{0,n}\\
\mathbf{\tilde{Y}}^{0,n}
\end{array}\right)\\
 & = & -\frac{1}{2}\textrm{Tr}\mathbf{U}_{ph,frac}\\
 & = & -\frac{1}{2}\sum_{ij}\left\langle ij\right|U_{ph}\left|ij\right\rangle _{frac}\\
 & = & -\frac{1}{2}\sum_{ij}\sqrt{(1-n_{i})n_{j}}\left\langle ij\right|U_{ph}\left|ij\right\rangle \sqrt{(1-n_{i})n_{j}}
\end{eqnarray*}
using
\[
\int_{-\infty}^{\infty}dE\textrm{Im}\mathbf{\Pi}^{0,N+\delta}(E)=-\pi\left[\left(\begin{array}{c}
\mathbf{\tilde{X}}^{0,n}\\
\mathbf{\tilde{Y}}^{0,n}
\end{array}\right)(\mathbf{\tilde{X}}^{0,n\dagger}\mathbf{\tilde{Y}}^{0,n\dagger})+\left(\left(\begin{array}{c}
\mathbf{\tilde{Y}}^{0,n}\\
\mathbf{\tilde{X}}^{0,n}
\end{array}\right)(\mathbf{\tilde{Y}}^{0,n\dagger}\mathbf{\tilde{X}}^{0,n\dagger})\right)^{*}\right]
\]

Finally

\begin{eqnarray*}
 &  & \frac{1}{4\pi}\int_{-\infty}^{\infty}dE\textrm{Im}\textrm{Tr}\left[\ln(\mathbf{1}-\mathbf{\Pi}^{0}(E)\mathbf{U}_{ph})+\mathbf{\Pi}^{0}(E)\mathbf{U}_{ph}\right]\\
 & = & \frac{1}{2}\sum_{n>0}\varepsilon_{n}^{\pi}-\frac{1}{2}\sum_{a\in\mathrm{particle},j\in\mathrm{hole}}\left(\varepsilon_{a}-\varepsilon_{j}\right)-\frac{1}{2}\sum_{ij}\left\langle ij\right|U_{ph}\left|ij\right\rangle _{f}\\
 & = & \frac{1}{2}\sum_{n>0}\varepsilon_{n}^{\pi}-\frac{1}{2}\textrm{Tr}\mathbf{A},
\end{eqnarray*}
which is the RPA correlation energy, Eq. (\ref{eq:E_integ}).
\end{document}